\documentclass[prd,onecolumn,nofootinbib]{revtex4}

\usepackage{subfigure}
\usepackage{amssymb}
\usepackage{epsfig}
\usepackage{float}
\usepackage{caption}
\usepackage{color}
\usepackage{upgreek}

\newcommand{\be}{\begin{equation}}
\newcommand{\ee}{\end{equation}}
\newcommand{\bea}{\begin{eqnarray}}
\newcommand{\eea}{\end{eqnarray}}

\input{tcilatex}

\begin{document}

\title{Perturbation theory for cosmologies with non-linear structure}

\author{Sophia R. Goldberg}
\email{s.r.goldberg@qmul.ac.uk}
\affiliation{School of Physics and Astronomy, Queen Mary University of London, Mile End Road, London E1 4NS, UK.}

\author{Christopher S. Gallagher}
\email{c.s.gallagher@qmul.ac.uk}
\affiliation{School of Physics and Astronomy, Queen Mary University of London, Mile End Road, London E1 4NS, UK.}

\author{Timothy Clifton}
\email{t.clifton@qmul.ac.uk}
\affiliation{School of Physics and Astronomy, Queen Mary University of London, Mile End Road, London E1 4NS, UK.}

\bibliographystyle{plain}

\begin{abstract}

The next generation of cosmological surveys will operate over unprecedented scales, and will therefore provide exciting new opportunities for testing general relativity. The standard method for modelling the structures that these surveys will observe is to use cosmological perturbation theory for linear structures on horizon-sized scales, and Newtonian gravity for non-linear structures on much smaller scales. We propose a two-parameter formalism that generalizes this approach, thereby allowing interactions between large and small scales to be studied in a self-consistent and well-defined way. This uses both post-Newtonian gravity and cosmological perturbation theory, and can be used to model realistic cosmological scenarios including matter, radiation and a cosmological constant. We find that the resulting field equations can be written as a hierarchical set of perturbation equations. At leading-order, these equations allow us to recover a standard set of Friedmann equations, as well as a Newton-Poisson equation for the inhomogeneous part of the Newtonian energy density in an expanding background. For the perturbations in the large-scale cosmology, however, we find that the field equations are sourced by both non-linear and mode-mixing terms, due to the existence of small-scale structures. These extra terms should be expected to give rise to new gravitational effects, through the mixing of gravitational modes on small and large scales -- effects that are beyond the scope of standard linear cosmological perturbation theory. We expect our formalism to be useful for accurately modelling gravitational physics in universes that contain non-linear structures, and for investigating the effects of non-linear gravity in the era of ultra-large-scale surveys.

\end{abstract}

\maketitle

\section{Introduction} \label{intro}

There is hope that the next generation of astronomical surveys \cite{SKA, euclid, LSST}, which will collect data on scales comparable to the cosmological horizon, will have sufficient precision to provide a new testing ground for non-linear relativistic gravity \cite{ Kitching:2015fra, Laureijs:2011gra, Amendola:2016saw, Abell:2009aa}. This is a particularly exciting prospect as, to date, non-linear gravitational effects have only been observed in the solar system \cite{GPB, Bertotti:2003rm, Everitt:2011hp, Biswas:2008cw}, binary pulsar systems \cite{Taylor:1989sw, Will:2014kxa}, and the newly discovered binary black hole mergers observed using LIGO \cite{PhysRevLett.116.061102, PhysRevLett.116.241103, PhysRevLett.118.221101}. The observation of corresponding effects in cosmology would allow general relativity to be investigated on entirely new length and time scales, as well as in a totally different physical environment. There is no telling what this might reveal about Einstein's theory.

To develop a mathematical formalism for investigating the non-linear properties of gravity in cosmology is, however, a highly non-trivial task. There is now a substantial literature dedicated to developing different approaches to modelling non-linear gravitational physics in cosmology. The most common approach is a direct implementation of second-order cosmological perturbation theory \cite{Malik:2003mv, Huston:2009ac, Nakamura:2006rk}, which allows relativistic gravitational perturbations around a homogeneous and isotropic background to be modelled in the presence of linear density contrasts. Other approaches, however, have started to import techniques from post-Newtonian gravity \cite{bruni, RDformalism, RDnature}, where gravitational fields are assumed to be slowly varying and where non-linear density contrasts can be consistently modelled.

These two different approaches are the obvious non-linear extensions of the standard method that has been so successful when considering linear gravity in cosmology; using cosmological perturbation theory to model linear structures on scales above the homogeneity scale ($\gtrsim 100$Mpc), and Newtonian gravity to model non-linear structures on smaller scales. This looks very natural at linear order in the gravitational fields, partly because the equations of Newtonian gravity can be recovered as the quasi-static limit of cosmological perturbation theory (when the gravitational fields slowly vary in time). Indeed, this pleasing feature extends to non-linear gravitational fields in cosmology \cite{bruni}. 

However, if one wants to consider non-linear gravity in a universe that {\it simultaneously} contains linear structures on large scales {\it and} non-linear structures on small scales, then one must adopt a more sophisticated approach. This is because, when considering the quasi-static limit, the terms that have been relegated to higher-order can no longer be entirely forgotten; they can and should be expected to appear in the next-to-leading-order gravitational field equations. This could be at second-order on small scales, but could in principle be at what is usually thought of as first-order on large scales. The question of how to construct a perturbative expansion that can systematically perform the required re-ordering, and produce a self-consistent and well-motivated set of field equations on all scales, is the purpose of this paper.

This work is a development of the two-parameter perturbative expansion proposed and investigated in a previous paper \cite{SRGTCKM}. We extend and apply this work to the case of realistic cosmological models that contain relativistic fluids with barotropic equations of state, as well as a cosmological constant, $\Lambda$, and non-relativistic dust-like matter that can be used to model dark matter and baryons. The result is a set of equations that can be used to calculate the effect of small-scale structure on the leading-order perturbations on large scales. These equations contain terms that are quadratic in short-scale potentials and can be written as an effective fluid \cite{Baumann:2010tm}, as well as mode-mixing terms\footnote{Note that we use ``mode-mixing'' to describe the coupling of scalar, vector and tensor perturbations. We will use ``scale-mixing'' to refer to the coupling of large-scale and short-scale perturbations due to quadratic terms.} that couple scalar, vector and tensor perturbations in the large-scale cosmology. Both of these two types of terms offer exciting possibilities for testing non-linear gravity with upcoming surveys.

In Section \ref{sec:pfe} we give the final field equations for the two-parameter perturbative expansion constructed during this paper, written in their effective fluid form. In Section \ref{sec:ThFr} we discuss the construction of the two-parameter perturbation expansion. Section \ref{Gchoice} contains a study of infinitesimal gauge transformations in the context of two-parameter expansions, as well as the construction of gauge-invariant variables. The application of these variables to the field equations is then discussed in Section \ref{sec:fe}.  We use Greek indices to represent spacetime coordinates, and Latin indices for spatial coordinates. Dots refer to differentiation with respect to cosmic time $\dot{\varphi} \equiv \frac{\partial \varphi}{\partial t}$, and dashes refer to differentiation with respect to conformal time $\varphi' \equiv \frac{\partial \varphi}{\partial \tau}$. Single spatial derivatives are given by $\varphi_{,i} \equiv \partial_i \varphi \equiv \nabla \varphi$, and $\nabla^2$ refers to the Laplacian associated with spatial partial derivatives with respect to the comoving coordinates. Throughout, we use units in which $G=c=1$. 

\section{Perturbed Field Equations}
\label{sec:pfe}

In this section we will present the perturbed field equations that result from simultaneously considering linear structures on large scales, and non-linear single-stream structures on small scales. The quantities that appear in these equations will then be explained further in the sections that follow, with technical details reserved for the appendices. We hope this will allow the busy reader to see the most physically interesting aspects of this work without first having to undergo the lengthy mathematical considerations that went into their derivation. It will, however, require the reader to temporarily suspend their satisfaction of knowing exactly what all the quantities involved represent, as we then move on to more carefully lay out their specific technical definitions and properties. The multi-stream generalization should follow straightforwardly. 

The field equations we present will be expressed in terms of the following set of gauge-invariant gravitational fields:
\begin{equation}
\{ {\rm U}, \upphi , \uppsi , {\rm S}_i , {\rm h}_{ij} \} \, , 
\end{equation}
as well as a corresponding set of gauge-invariant matter perturbations:
\begin{equation}
\{ \updelta \uprho_{\rm N}, \updelta \uprho, \updelta {\rm p}, {\rm v}_{{\rm N} i}, {\rm v}_i  \} \, ,
\end{equation}
where $\uprho$, ${\rm p}$ and ${\rm v}_i $ correspond to the energy-density, pressure and peculiar velocity, respectively. These quantities all represent fluctuations about a spatially-flat FLRW geometry, which in a particular choice of coordinates can be written as
\begin{equation} \label{FLRW1}
ds^2 = a^2(\tau) \Big[ -(1 + 2 {\rm U}+ 2 \upphi) d\tau^2 + \big( (1-2 {\rm U} - 2 \uppsi)\delta_{ij} + 2 {{\rm h}}_{ij} \big) dx^i dx^j   -2 {\rm S}_i d\tau dx^i \Big] \, ,
\end{equation}
where $a$ is the scale factor and $\tau$ is a conformal time coordinate. In what follows we will also require the Hubble rate in conformal time, which we write as $\mathcal{H} \equiv a'/a$.

There are a couple of things that require some explanation before we can proceed further. First, $\rm U$ has been removed from the scalar potentials $\upphi$ and $\uppsi$ because we intend it to correspond to the Newtonian gravitational potential. A precise explanation of what we mean by this will come in the following sections. For now it sufficient to understand ${\rm U}$ as the leading-order part of the gravitational field produced by non-linear density contrasts. The potentials $\upphi$ and $\uppsi$, as well as ${\rm h}_{ij}$ and ${\rm S}_i$, contain information about both the large-scale cosmological potentials and the small-scale post-Newtonian potentials. Again, the precise meaning of this will become clearer in subsequent sections. Likewise, the Newtonian density contrast is $\updelta \uprho_{\rm N}$, and the cosmological \textit{and} post-Newtonian density contrast is given by $\updelta \uprho$. The former of these is allowed to be arbitrarily large, while the latter is required to be small (again, to be made precise later on). Similar comments apply to ${\rm v}_{{\rm N}i}$ and ${\rm v}_i $.

After simultaneously expanding the field equations in post-Newtonian and cosmological perturbation theories we find the leading-order parts are given by the effective Friedmann equations
\begin{equation} \label{fried}
\mathcal{H}^2 = \frac{8\pi a^2}{3} \bar{\uprho} + \frac{1}{3}\Lambda a^2 +\mathcal{O}(\eta^4)
\qquad {\rm and} \qquad
\mathcal{H}'  = -\frac{4\pi a^2}{3}(\bar{\uprho} + 3 \bar{{\rm p}}) + \frac{1}{3}\Lambda a^2 +\mathcal{O}(\eta^4) \, ,
\end{equation}
where $\bar{\uprho}=\bar{\uprho}_{\rm M}+\bar{\uprho}_{\rm R}$ and $\bar{{\rm p}}=\bar{{\rm p}}_{\rm R}$ are the leading order parts of the spatial averages of the energy density and pressure, respectively. They have both radiation ($\bar{\uprho}_{\rm R}$ and $\bar{{\rm p}}_{\rm R}$),  and dark and baryonic matter ($\bar{\uprho}_{\rm M}$) contributions. The Newtonian gravitational field equation occurs at the same order in our expansion, and is given by
\begin{equation} \label{multipoisson}
\nabla^2 {\rm U} = 4\pi a^2 \updelta \uprho_{\rm N} +\mathcal{O}(\eta^4) \, ,
\end{equation}
where $\eta$ is the expansion parameter for the post-Newtonian expansion. In this case it is used to characterise the size of structure on scales of order the homogeneity scale; the largest-scale at which the post-Newtonian expansion can sensibly be performed. The reader may note that only dark matter and baryonic matter contribute to $\updelta \uprho_{\rm N}$, and not radiation.

Subsequent orders of the perturbation expansion in the field equations yield the following two equations for the scalar part of the gravitational field:
\begin{eqnarray} \label{upphi}
\frac{1}{3} \nabla^2 {\upphi} + \mathcal{H}\upphi' + \mathcal{H}\uppsi' + \uppsi'' + 2\mathcal{H}'\upphi   
&=& \frac{4\pi a^2}{3} \big( \updelta \uprho + \updelta \uprho_{\rm eff} + 3 \updelta {\rm p} +  3 \updelta {\rm p}_{\rm eff} \big) 
+ \frac{2}{3} (\mathcal{D}^{ij} {\rm U}) {\rm h}_{ij} - \frac{8 \pi a^2}{3} \updelta \uprho_{\rm N} (\uppsi - \upphi) +\mathcal{O}(\eta^5) \qquad
\end{eqnarray}
and
\begin{eqnarray}
\frac{1}{3} \nabla^2 \uppsi - \mathcal{H}\uppsi' - \mathcal{H}^2\upphi  
&=& \frac{4\pi a^2}{3} \big( \updelta \uprho + \updelta \uprho_{\rm eff} \big) 
+ \frac{1}{3} (\mathcal{D}^{ij} {\rm U}) {\rm h}_{ij} - \frac{16 \pi a^2}{3} \updelta {\uprho}_N \uppsi +\mathcal{O}(\eta^5) \, ,
\end{eqnarray}
where $D_{ij} \varphi \equiv \varphi_{,(ij)} - \frac{1}{3} \delta_{ij} \nabla^2 \varphi$ is the trace-free second derivative operator on any field $\varphi$, and where perturbations in radiation, and dark and baryonic matter contribute to both $\updelta \uprho$ and $\updelta {\rm p}$. The reader will note that these equations contain extra terms when compared to standard cosmological perturbation theory. First, there are effective energy density and pressure terms, $\updelta \uprho_{\rm eff}$ and $\updelta {\rm p}_{\rm eff}$. These are solely due to the presence of non-linear structures on small scales, and are given explicitly in Eqs. (\ref{deltarhoeff}) and (\ref{deltapeff}), below. In other words, by writing the field equations in an effective fluid description, one can clearly identify that small-scale non-linearities lead to, amongst other things, an effective pressure on large-scales. Second, in the above equations, the potential ${\rm U}$ couples to ${\rm h}_{ij}$ and there are extra source terms on the right-hand-side of these equations that are linear in $\upphi$ and $\uppsi$. These interaction terms do not exist in standard cosmological perturbation theory and vanish in the limit in which non-linear small-scale structures vanish. In general, the interaction terms should be expected to produce mode-mixing between scalar, vector and tensor parts of the gravitational field on cosmological scales and coupling between different Fourier modes in Fourier-space.

The remaining parts of the gravitational field are the vector and tensor modes. For the vectors we find that we can write the following single equation to describe ${\rm S}_i$, accurate up to order $\mathcal{O}(\eta^5)$:
\begin{eqnarray} \label{vec4}
\nabla^2 {\rm S}_i +4\partial_i \big( \uppsi'  + \mathcal{H}\upphi \big) +16\pi a^2 \big(\bar{\uprho}  + \bar{{\rm p}} + \updelta \uprho_{\rm N} \big) ({\rm v}_i  - {\rm S}_i)
&=& - 16\pi a^2 Q_i^{{\rm eff}} - 8\pi a^2  \updelta \uprho_{\rm N}  {\rm S}_i -2  ( \partial_j \partial_i {\rm U}) {\rm S}^j +\mathcal{O}(\eta^5) \, .
\end{eqnarray}
We can take the leading-order part of this equation, at $\mathcal{O}(\eta^3)$, and write it as the following simple Poisson equation
\begin{eqnarray} \label{vec3}
\nabla^2 {\rm S}_i  +4 \partial_i ({\rm U}' + \mathcal{H}{\rm U}) +16\pi a^2 \big(\bar{\uprho}  + \bar{{\rm p}} \big) {\rm v}_{{\rm N}i} &=&    -16\pi a^2  \updelta \uprho_{\rm N} {\rm v}_{{\rm N}i} +\mathcal{O}(\eta^4) \, .
\end{eqnarray}
The leading-order part of the vector gravitational field, given by the solution to Eq. (\ref{vec3}), is only sourced by small-scale quantities,  and is a hundred times greater than might naively be expected from cosmological perturbation theory. This is the equation that was identified in the post-Friedmann approach of \cite{bruni}, and solved for numerically in \cite{bruniNbody}. For the full vector equation (\ref{vec4}), accurate up to $\mathcal{O}(\eta^5)$, it can be seen that there exists sources on both small and large scales and mode-mixing  (which is missing from \cite{Adamek:2015hqa, Adamek:2014xba}). This equation has an effective energy flux, $Q_i^{{\rm eff}}$, which is due to small scale potentials. It also has extra source terms on the right-hand-side that are linear in $S_i$. Both of these vanish when small-scale structures are absent. The explicit expression for $Q_i^{{\rm eff}}$ is given in Eq. (\ref{Qieff}), below, along with the other effective fluid quantities.

The final field equations we require, in order to complete our set to the desired accuracy, is given as follows:
\begin{equation} \label{tracefreecosmo}
\nabla^2 {\rm h}_{ij} - {\rm h}''_{ij} - 2\mathcal{H} {\rm h}'_{ij} 
+\mathcal{D}_{ij}(\upphi - \uppsi) - 2\mathcal{H}\partial_{(j}{\rm S}_{i)}  - \partial_{(j}{\rm S}'_{i)}
= -8\pi a^2 \Pi_{ij}^{\rm eff} - 8 \pi a^2 \updelta \uprho_{\rm N} {\rm h}_{ij} + 4 (\partial^k \partial_{\langle i}{\rm U}) {\rm h}_{j\rangle k} + 2(\mathcal{D}_{ij} {\rm U}) (\upphi + \uppsi) +\mathcal{O}(\eta^5) \, ,
\end{equation}
where angle brackets around indices indicate a symmetric and trace-free operation has been used, so that $\mathcal{T}_{\langle ij \rangle} \equiv \mathcal{T}_{(ij)} - \frac{1}{3} \delta_{ij} \mathcal{T}_{kk}$ for any field $\mathcal{T}_{ij}$. This equation can be used to determine the tensor part of the gravitational field, ${\rm h}_{ij}$. It also has an effective fluid source, $\Pi_{ij}^{\rm eff}$, which this time acts as an effective anisotropic stress and is formed from the quadratic contractions of the lower-order small-scale potentials, see Eq. (\ref{Piijeff}). Again, the non-linear structure on small scales couples the large-scale scalar and tensor parts of the cosmological gravitational fields, and again we have additional terms on the right-hand-side that are linear in ${\rm h}_{ij}$, resulting in mode-mixing.

As promised, the effective fluid quantities in the perturbation equations above are given as follows:
\begin{eqnarray}
\updelta \uprho _{\rm eff} &=&  (\bar{\uprho}  + \bar{{\rm p}} + \updelta \uprho_{\rm N})({\rm v}_{\rm N})^2
  - \frac{1}{\pi a^2} {\rm U}\nabla^2 {\rm U} + \frac{3}{4\pi a^2}\Big(\mathcal{H}^2 {\rm U} + \mathcal{H}{\rm U}' - \frac{1}{2}(\nabla {\rm U} )^2 \Big) \label{deltarhoeff}
\\[10pt]
\updelta {\rm p}_{\rm eff} &=&  \frac{1}{3}(\bar{\uprho}  + \bar{{\rm p}} +\updelta \uprho_{\rm N})({\rm v}_{\rm N})^2 - \frac{1}{4\pi a^2} \Big({\rm U}'' + 3\mathcal{H} {\rm U}' - \frac{7}{6}(\nabla {\rm U})^2 +   a^2 {\rm U}  (\Lambda - 8\pi \bar{p})\Big) + \frac{1}{3\pi a^2} {\rm U} \nabla^2 {\rm U} \label{deltapeff}
\\[10pt]
Q_i^{{\rm eff}} &=& \big(\bar{\uprho}  + \bar{{\rm p}}  + \updelta \uprho_{\rm N} \big) {\rm v}_{{\rm N}i} + \frac{1}{4 \pi a^2} \partial_i ({\rm U}' + \mathcal{H}{\rm U})   \label{Qieff}
\\[10pt]
\Pi_{ij}^{\rm eff} &=& (\bar{\uprho}  + \bar{{\rm p}}+\updelta \uprho_{\rm N}) {\rm v}_{{\rm N}\langle i} {\rm v}_{{\rm N} j\rangle} - \frac{1}{4\pi a^2} \partial_{\langle i}{\rm U} \partial_{j \rangle} {\rm U}  - \frac{1}{2\pi a^2} {\rm U} \mathcal{D}_{ij} {\rm U} \, . \label{Piijeff}
\end{eqnarray}
It can be seen that each of these quantities was constructed only from variables that correspond to small-scale gravitational fields, or background quantities (which will be shown later to be calculated from the average of small-scale quantities). We therefore have a hierarchy of equations that can be solved order-by-order: first, the Friedmann and Newtonian equations (\ref{fried}) and (\ref{multipoisson}), and then the equations that contain large-scale perturbations (\ref{upphi})-(\ref{tracefreecosmo}). The former of these sets are already calculated routinely in modern cosmological N-body simulations. The latter are modified versions of the usual cosmological perturbation equations on large scales, and can be used to find post-Newtonian equations on small scales (as recently solved for numerically in \cite{RDgevolution, RDnature, Adamek:2015hqa, Adamek:2014xba}). Finally, note that the above effective quantities, in Eqs. (\ref{deltarhoeff})-(\ref{Piijeff}), contain terms that would normally only be included in second or third order in cosmological perturbation theory.  In particular, the term $\updelta \uprho_{\rm N} {\rm v}_{{\rm N}\langle i} {\rm v}_{{\rm N} j\rangle}$ in Eq. (\ref{Piijeff}) would appear at third order in standard perturbation theory, but here should be expected to source a gravitational ``slip'' in the leading-order part of the large-scale physics. Our approach can be compared to the effective fluid approach studied previously in \cite{Baumann:2010tm, Senatore},  as well as the large and small wavelength split used in \cite{GreenWald1, GreenWald2}. 

As a final comment, before moving on to explain the origin of these equations and give detailed explanations of the gauge invariant quantities involved, we note that the usual trick of separating equations like (\ref{vec4}) and (\ref{tracefreecosmo}) into scalar, vector and tensor parts is much more difficult to apply here. This is due to the fact terms like $(\mathcal{D}_{ij} {\rm U}) (\upphi + \uppsi)$ do not have scalar, vector and tensor parts that are easy to identify. This term, for example, is a scalar multiplied by a tensor, and in general should be expected to contain scalar, vector and tensor parts. This does not mean that such a separation is impossible -- indeed we very much expect it to be possible. It just means that the resulting equations are very messy to write down, which is the reason why we have chosen to present these equations without such a decomposition. The reader should also be warned that manipulation of these equations is considerably more difficult than in either cosmological perturbation theory or standard post-Newtonian theory. This is due to different derivative operators changing the order to the terms they operate on in different ways. This will be made clearer in the sections that follow, and expanded upon in more detail in a subsequent publication \cite{gallagher}.

\section{Perturbative frameworks}
\label{sec:ThFr}

In order to explain the origin of the equations presented in Section \ref{sec:pfe} we first need to outline the perturbative expansions used in their derivation. These are the post-Newtonian expansion and cosmological perturbation theory, which will be used to describe perturbations on small and large scales, respectively. For the former of these expansions we use the expansion parameter $\eta \ll 1$, while for the latter we use $\epsilon \ll 1$. We assume that any field $Q$ can be expanded in both $\epsilon$ and $\eta$ as follows:
\be
Q = \sum\limits_{n,m} \frac{1}{n!m!}Q^{(n,m)} \, ,
\ee
where $Q^{(n,m)}$ is a quantity of order $\mathcal{O}(\epsilon^n \eta^m)$. The characteristic length scales on which post-Newtonian and cosmological perturbations exist and vary on will be labelled $L_N$ and $L_C$, respectively. Let us now briefly outline the essential features of these two different expansion schemes, before considering how they can be performed simultaneously.

\subsection{Post-Newtonian theory}
\label{PostNBk}

Post-Newtonian gravity is a slow-motion and weak-field expansion of the field equations, usually applied to systems that are much larger than their Schwarzschild radius but also much smaller than the Hubble radius \cite{will}. Characteristic dimensionless velocities on such scales are usually small, from which it follows that time derivatives of fields are also small in comparison to spatial derivatives:
\be \label{smallderivs}
\dot{\varphi} \ll  \vert \nabla \varphi \vert  \sim \frac{\varphi}{L_N} \, ,
\ee
where $L_N$ is the spatial length scale over which these perturbations are expected to vary, and $\varphi$ represents metric potentials. The perturbative order-of-smallness of metric potentials is given by noting that the magnitude of peculiar velocities are small, $| v^{(0,1)i} | \sim \eta$, and assigning terms with time derivatives an extra order-of-smallness in $\eta$, compared to spatial derivatives. The magnitude of the gravitational potentials, in terms of $\eta$, can then be determined through the geodesic equations for freely falling particles and the field equations. This process can also be used to assign orders of smallness to the sources of stress-energy \cite{poisson, will}.

At this stage orders-of-smallness in $\eta$ can be assigned to all matter fields and gravitational potentials. The perturbed energy density and pressure for a perfect fluid are given by the expansions $\rho = \rho^{(0,2)} + \frac{1}{2}\rho^{(0,4)} + \ldots$ and $p = p^{(0,4)} + \ldots$, respectively, where ellipses denote higher-order terms. The quantity $\rho^{(0,2)}$ is the Newtonian energy density, while $\rho^{(0,4)}$ and $p^{(0,4)}$ are the post-Newtonian contributions to energy density and pressure. The reader may note that there is no pressure term at $\mathcal{O}(\eta^2)$. This is no accident -- if such a term were to be included then the stress-energy conservation equations would require it to be spatially homogeneous. This means that barotropic fluids with $p= w \rho$ and $w \neq 0$ do not fit into post-Newtonian gravity in a natural way, unless they are diffuse enough to be considered post-Newtonian (i.e. to occur only at $\mathcal{O}(\eta^4)$ or above).

Similarly, order-of-smallness can be assigned to the metric components. The minimal set of metric perturbations required to consistently expand a time-dependent background metric, $g^{(0,0)}_{\mu \nu} (t)$, using the post-Newtonian expansion, is therefore given by
\bea \nonumber
\delta g_{00} &=& \delta g_{00}^{(0,2)} (t,{\bf x}) + {\textstyle\frac{1}{2}} \delta g_{00}^{(0,4)} (t,{\bf x}) \ldots , \qquad
\delta g_{ij} = \delta g_{ij}^{(0,2)} (t,{\bf x}) + \ldots , \qquad {\rm and} \qquad  
\delta g_{0i} = \delta g_{0i}^{(0,3)} (t,{\bf x}) + \ldots\, \, .
\eea
It is important to note that the post-Newtonian formalism outlined above was formulated to include matter, but not radiation or a cosmological constant. This will be discussed further in Section \ref{radL} below, in the context of incorporating post-Newtonian expansions into cosmology. For further details about post-Newtonian expansions, the reader is referred to the textbooks by Will \cite{will}, and Poisson and Will \cite{poisson}.

\subsection{Cosmological perturbation theory} 
\label{CPTBk}

Cosmological perturbation theory is a weak-field expansion about the spatially homogeneous and isotropic class of Friedmann-Lema\^{i}tre-Robertson-Walker (FLRW) geometries:
\be
\label{flrw}
ds^2 = -dt^2 + a^2(t) \left( \frac{dr^2}{1-k r^2} +r^2 (d\theta^2+ \sin^2 \theta d\phi^2) \right) \, ,
\ee
where $a(t)$ is the scale factor and $k$ is the spatial curvature (we take $k=0$ for the rest of this paper). All perturbations to all matter and gravitational fields in this perturbative expansion are taken to occur at the same order:
\be \label{epsilon}
\epsilon \sim  |v^{(1,0)i} | \sim L_C^2 \, \delta \rho^{(1,0)} \sim L_C^2 \, \delta p^{(1,0)} \sim \delta g_{\mu \nu}^{(1,0)}   \, ,
\ee
where $L_C$ is the characteristic length scale gravitational fields vary on, and is necessary above to compare the dimensionless expansion parameter, peculiar velocity and gravitational potentials to dimensionful quantities like the perturbed energy density and pressure. Cosmological perturbation theory also assumes that time and space derivatives do not change the order-of-smallness of a quantity, so we have
\be 
\dot{\varphi} \sim \vert \nabla \varphi \vert \sim \frac{\varphi}{L_C} \, .
\ee
This result occurs because the perturbations can be as large as, or larger than, the horizon, and is in stark contrast with the post-Newtonian expansion outlined previously. It means that orders at which terms appear in the two expansions cannot in general be expected to be comparable, a difference that changes the character of the resulting field equations.

For our present purpose it is important to note that standard cosmological perturbation theory is designed in such a way that it can be applied to many different epochs in the Universe. Specifically, given a fluid with equation of state $p =w\rho$ then both the background and perturbed equations can be written down straightforwardly. This allows the theory to be applied to both the radiation-dominated and matter-dominated stages of the Universe's evolution, as well as to the current cosmological-constant-dominated epoch. This is a versatility that is absent from the standard approach in post-Newtonian gravity, as radiation and $\Lambda$ are completely negligible for the study of gravity in the Solar System, binary pulsars, and other such very-small-scale astrophysical environments. If one wants to apply such expansions to super-clusters in a cosmological context, however, then more care may be required. For further explanation of cosmological perturbation theory, the reader is referred to the review by Malik and Wands \cite{malik}.

\subsection{Two-parameter perturbation theory} 
\label{pert2para}

In reality, both post-Newtonian and cosmological perturbations should be expected to exist in any realistic model of the Universe \cite{Planelles:2014zaa}. In this section we review and extend the two-parameter framework developed in \cite{SRGTCKM} that simultaneously performs a perturbative expansion in both sectors. In the next section we will extend this formalism to include radiation and a cosmological constant. Our two-parameter expansion in both $\epsilon$ and $\eta$ will be constructed around an FLRW geometry, corresponding to the line-element in Eq. (\ref{flrw}). This is the standard background for cosmological perturbation theory, but is so far little used for post-Newtonian gravity (see however \cite{bruni,  RDformalism}). Nevertheless, it can be shown that both expansions can be performed in such a background in an entirely self-consistent and well-posed way \cite{cwoa,sanghai1,SRGTCKM}. 

The first step in doing this is to expand the total energy density and pressure in both $\epsilon$ and $\eta$:
\bea \label{rhom}
\rho &=&  \rho^{(0,0)} + \rho^{(0,2)} + \rho^{(1,0)} + \rho^{(1,1)} + \rho^{(1,2)} + {\textstyle\frac{1}{2}}\rho^{(0,4)} + \ldots \\ \label{rhor}
p &=& p^{(0,0)} + p^{(1,0)} + p^{(1,2)} + {\textstyle\frac{1}{2}}p^{(0,4)} + \ldots \, .
\eea
The terms $\rho^{(0,0)}$ and $p^{(0,0)}$ can be considered as the background energy density and pressure, as they are not perturbed in either $\epsilon$ or $\eta$. All other terms correspond to perturbations at the order indicated by the superscript, but we have neglected to include $\delta$ symbols before them to keep the notation as compact as possible. To be even more precise, the orders-of-magnitude of these perturbed quantities are given by
\begin{eqnarray} \label{rhomags}
&& \rho^{(0,0)} \sim \frac{1}{L_C^2} \, , \qquad \rho^{(n,0)} \sim \frac{\epsilon^n}{L_C^2} \, , \qquad \rho^{(0,m)} \sim \frac{\eta^m}{L_N^2} \qquad \mathrm{and} \qquad \rho^{(n,m)} \sim \frac{\epsilon^{n}\eta^m}{L_N^2} \, ,
\end{eqnarray}
where $\{ m, n\} \in \mathbb{N}^{+}$, and again $L_C$ and $L_N$ are the characteristic length scales of the cosmological and post-Newtonian sytems, respectively. A similar expression holds for the expansion of $p$. The length scales are necessary in the denominators of these expressions, as $\rho$ is a quantity with dimension $L^{-2}$, and because it only makes sense to compare the magnitude of quantities with the same dimensions. The first thing to notice about Eq. (\ref{rhom}) is that the mixed-order terms do not always appear at the same order as the product of post-Newtonian and cosmological terms (i.e. we have included $\rho^{(1,1)}$, even though there is no $\mathcal{O}(\eta)$ term in the post-Newtonian expansion). The reason for this is that such terms are necessarily generated by arbitrary gauge transformations, and so cannot be assumed to vanish in general, even if they are assumed to do so in one particular coordinate system. Secondly, we have included a background energy density, which was absent in Ref. \cite{SRGTCKM}. This will be useful for including radiation.

We also need to expand the metric in both $\epsilon$ and $\eta$, which we do as follows:
\bea
g_{00} &=& g_{00}^{(0,0)} + g_{00}^{(0,2)}  + g_{00}^{(1,0)} + g_{00}^{(1,1)}+ g_{00}^{(1,2)} + {\textstyle\frac{1}{2}}g_{00}^{(0,4)} + \ldots  \label{g00}\\
&=& -1 + h_{00}^{(0,2)}  + h_{00}^{(1,0)} + h_{00}^{(1,1)}+ h_{00}^{(1,2)} + {\textstyle\frac{1}{2}}h_{00}^{(0,4)} + \ldots \nonumber  \\[10pt]
g_{ij} &=& g_{ij}^{(0,0)} + g_{ij}^{(0,2)}  + g_{ij}^{(1,0)}+ g_{ij}^{(1,1)} + g_{ij}^{(1,2)}  + {\textstyle\frac{1}{2}}g_{ij}^{(0,4)} + \ldots \label{gij}\\
&=& a^2 \left( \delta_{ij} + h_{ij}^{(0,2)}  +h_{ij}^{(1,0)} + h_{ij}^{(1,1)} + h_{ij}^{(1,2)}    +{\textstyle\frac{1}{2}}h_{ij}^{(0,4)} \right)  + \ldots \nonumber \\ [10pt]
g_{0i} &=& g_{0i}^{(1,0)}  + g_{0i}^{(0,3)} + g_{0i}^{(1,2)}+ \ldots  \label{g0i} \\
&=& a \left( h_{0i}^{(1,0)}  + h_{0i}^{(0,3)}+ h_{0i}^{(1,2)}  \right)+ \ldots \, ,  \nonumber
\eea
where in the second line of each of these equations we have chosen our background metric $g^{(0,0)}_{\mu \nu}$ to be the flat FLRW metric from Eq. (\ref{flrw}), and simultaneously defined the perturbations $h_{\mu \nu}$. The orders of magnitude of each of the perturbations to each of the components of this metric are the minimal set required to self-consistently account for the gravitational fields of the two-parameter perturbed perfect fluid discussed above, in any arbitrary coordinate system. We find that the inclusion of radiation and a cosmological constant does not require the introduction of any new metric potentials at any new order, so the form of Eqs. (\ref{g00})-(\ref{g0i}) is the same as in Ref. \cite{SRGTCKM}. 

The final ingredient of the field equations that must be perturbed is the peculiar velocity, $v^i$. This is split into post-Newtonian and cosmological parts such that 
\bea \label{vee}
v^i &=& v^{(0,1)i} + v^{(1,0)i} + \ldots \, ,
\eea
which leads to the following components of the reference four-velocity $u^{\mu}$:
\bea
u^0&=& 1 + \frac{1}{2}\left( h^{(0,2)}_{00} + h^{(1,0)}_{00} \right) + {\textstyle\frac{1}{2}}v^{(0,1)i}v^{(0,1)}_i + \ldots \; \; \; \; \; \; \; \label{uu0}\\[10pt]
u^i&=& \frac{1}{a}\left( v^{(0,1)i}+ v^{(1,0)i}\right) + \ldots \, ,\label{uui}
\eea
which are derived using the normalization condition $u^{\mu}u_{\mu}=-1$, and Eqs. (\ref{g00})-(\ref{g0i}). The components of the two-parameter perturbed energy-momentum tensor that arise from these equations are given in Appendix \ref{appendixlong}. The components of the Ricci tensor are unchanged from Ref. \cite{SRGTCKM}, and can be found in the appendix of that paper.

The reader should note that within the context of the two-parameter formalism, time derivatives are taken to add an extra order-of-smallness, $\eta$, compared to spatial derivatives whenever they act on an object that contains any non-zero perturbation in its post-Newtonian sector. So, for example, we take
$$
\dot{\rho}^{(0,2)} \sim \eta \, \vert \nabla \rho^{(0,2)} \vert \sim \frac{\eta^3}{L_N^3} \qquad {\rm and} \qquad
\dot{\rho}^{(1,1)} \sim \eta \, \vert \nabla \rho^{(1,1)} \vert \sim \frac{\epsilon \eta^2}{L_N^3} \qquad {\rm whilst} \qquad \dot{\rho}^{(1,0)} \sim \vert \nabla \rho^{(1,0)} \vert \sim \frac{\epsilon}{L_C^3} \, .
$$
As in Eq. (\ref{rhomags}), the purpose of this is to reflect the expectation that quantities perturbed in the post-Newtonian sector should be slowly varying in time and change over spatial length scales $L_N$, while quantities that are perturbed only in the cosmological sector should vary equally over both time and length scales $L_C$. This is explained further in \cite{SRGTCKM}.\\

\subsection{Including radiation and ${\mathbf \Lambda}$}
\label{radL}

Let us now consider how to add radiation and $\Lambda$ to our two-parameter expansion. For radiation this can be achieved by writing
\bea
\rho &=& \rho_M + \rho_R \, , \qquad p = p_M + p_R \qquad {\rm and } \qquad \rho\, v^i = \rho_M \, v^{i}_M + \rho_R \, v^{i}_R  \qquad {\it etc.} \, ,
\eea
where $\rho_M$ and $p_M$ are the energy density and pressure of non-relativistic matter, $\rho_R$ and $p_R$ are the energy density and pressure of radiation, and $v_M^i$ and $v^i_R$ are the peculiar velocities of the matter and radiation fluids. We then want to expand each of these new quantities in $\epsilon$ and $\eta$, which we do according to
\bea \label{rhoexpmat}
\rho_M  
&=&  \rho^{(0,2)}_M+ \rho^{(1,0)}_M  +\rho^{(1,1)}_M + \rho^{(1,2)}_M  + {\textstyle\frac{1}{2}}\rho^{(0,4)}_M + \ldots
\\ \label{pexp}
p_M  &=& p^{(1,0)}_M  +p^{(1,2)}_M + {\textstyle\frac{1}{2}}p^{(0,4)}_M+ \ldots \\ 
\label{rhoexprad}
\rho_R  &=&  \rho^{(0,0)}_R+ \rho^{(1,0)}_R  + \rho^{(1,2)}_R  + {\textstyle\frac{1}{2}}\rho^{(0,4)}_R + \ldots \\
p _R &=& p^{(0,0)}_R + p^{(1,0)}_R + p^{(1,2)}_R+ {\textstyle\frac{1}{2}}p^{(0,4)}_R + \ldots \, , 
\label{pexprad}
\eea
and
\bea \label{vexpmat}
v^i_M &=& v^{(0,1)i}_M + v^{(1,0)i}_M + \ldots  \, ,
 \qquad
v^i_R = v^{(0,1)i}_R + v^{(1,0)i}_R + \ldots \, .
\eea
These equations can, of course, be compared to Eqs. (\ref{rhom}), (\ref{rhor}) and (\ref{vee}) to read off values for the perturbations to the total energy density, pressure and $v^i$. They can also be seen to generalise those from Ref. \cite{SRGTCKM} by the inclusion of $\rho_R$ and $p_R$, as well as by the inclusion of an extra mixed-order term $p^{(1,2)}_M$ and a factor of $1/2$ in front of $p^{(0,4)}_M$. 

The reader will note that the expansions of the matter and radiation fluids have not been performed in an identical way: We have omitted (i) a time-dependent background-level contribution to the matter energy density and pressure, and (ii) a Newtonian-level contribution to the radiation energy density and pressure, so that
$$\rho^{(0,0)}_M=p^{(0,0)}_M=0  \qquad {\rm and} \qquad  \rho^{(0,2)}_R=p^{(0,2)}_R=0 \, .$$
The former of these is neglected because it corresponds to a term that could otherwise be as large as the Newtonian rest mass energy density $\rho^{(0,2)}_M$, which we consider to be highly unphysical. 

In the real universe there is no time-dependent background matter component to the energy density, $\rho^{(0,0)}_M(t)$. This is because the leading-order contribution to $\rho_{M}$ is in fact dominated by the (inhomogeneous) rest mass of galaxies, dust \textit{etc.}, which is exactly what $\rho^{(0,2)}_M(x^{\mu})$ corresponds to. 
Furthermore, $\rho^{(0,0)}_M$ would necessarily have to be a function of time only and because there is no discernible homogeneous fluid of non-relativistic matter with this magnitude in the real Universe\footnote{In fact, the existence of such a component corresponds to a breakdown of standard perturbation theory \cite{Rasanen}.}. The term $p^{(0,0)}_M$ could be neglected on similar grounds, but must also vanish because of the requirement $p \ll \rho$ in non-relativistic matter. 

Let us now consider the expansion of $\rho_R$ and $p_R$ given in Eqs. (\ref{rhoexprad}) and (\ref{pexprad}). For this purpose it is useful to consider the stress-energy conservation equation for the total stress-energy tensor $T_{\mu \nu}$:
\bea 
\nabla^{\mu}T_{\mu \nu} =\nabla^{\mu} (T_{M \mu \nu } + T_{R\mu \nu })= 0 \, , \label{conEq} 
\eea
where $T_{M \mu \nu }$ and $T_{R\mu \nu }$ are the matter and radiation contributions to the total stress-energy tensor, respectively. This implies $\nabla^{\mu} T_{M \mu \nu } =  Q_{\nu} $ and $\nabla^{\mu} T_{R \mu \nu } = - Q_{\nu} $, where $Q_{\nu} \neq 0$ for interacting fluids and $Q_{\nu}= 0$ for non-interacting fluids. In either case, the lowest-order part of Eq. (\ref{conEq}) is given by
\bea
\nabla p^{(0,0)}_R&=&0 \, , \label{p00timedep}
\eea
which implies $p^{(0,0)}_R=p^{(0,0)}_R(t)$ is a function of time only. If we now take $p_R = \frac{1}{3}\rho_R$, then this result implies that the leading-order part of the energy density in radiation must also be spatially homogeneous, such that $\rho^{(0,0)}_R=\rho^{(0,0)}_R(t)$. This is, in fact, exactly what is required for a background-level contribution to the energy density in an FLRW model.

A similar argument can now be used to understand why it would be inappropriate to include a term $\rho^{(0,2)}_R$ in Eq. (\ref{rhoexprad}). Such a term would imply the existence of $p^{(0,2)}_R$ which, again through the conservation equations, can be shown to be necessarily spatially homogeneous. Such a term would therefore be functionally degenerate with $\rho^{(0,0)}_R$, as they are both functions of time only, and would therefore show up in every conceivable set of equations in exactly the same way. We can therefore neglect both $\rho^{(0,2)}_R$ and $p^{(0,2)}_R$ without any loss of generality. Moreover, the term $\rho^{(0,2)}_R(t)$  would be Newtonian in size, and such a term would be highly unusual in normal post-Newtonian gravity. We therefore find that the lowest order at which inhomogeneous perturbations in radiation fit into our two-parameter expansion is at order $ \mathcal{O}(p^{(1,0)}_R) \sim \mathcal{O}(\epsilon L_C^{-2})$, which corresponds to a cosmological-scale perturbation.

The reader may also note that there is no term $\rho^{(1,1)}_R$  in Eq. (\ref{rhoexprad}), whereas there is a term $\rho^{(1,1)}_M$ in Eq. (\ref{rhoexpmat}). The $\rho^{(1,1)}_M$ is necessary because a term of the form $\rho^{(0,2)}_{M,i}\xi^{(1,0)i}$ is always generated  under a general infinitesimal gauge transformation \cite{SRGTCKM} (where $\xi^{(1,0)i}$ is a part of the gauge generator -- see Section \ref{Gchoice}). This implies there must in general exist a term $\rho^{(1,1)}_M$ in the expansion of $\rho_M$, because even if we artificially exclude it in one coordinate system, it will be generated in another. However, a similar argument does not apply to $\rho^{(1,1)}_R$, because the gauge transformation $\rho^{(0,0)}_{R}$ does not generate any terms of the same order as $\rho^{(1,1)}_R$. This can be seen to be true because $\rho^{(0,0)}_R$ is a function of time only, such that $\rho^{(0,0)}_{R,i}\xi^{(1,0)i}=0$. Of course, the same argument would apply to a term of the form $\rho^{(0,2)}_{R}$, if it had been included, as this term is also time dependent. This means that we can set $\rho^{(1,1)}_R=p^{(1,1)}_R=0$ in any coordinate system, and the same result will hold in any other coordinate system related by an infinitesimal gauge transformation.

Finally, let us consider the cosmological constant $\Lambda$. We assign an order of magnitude and dimensions to the cosmological constant in the following way:
\bea
\Lambda = \Lambda^{(0,0)} \sim \frac{1}{L_C^2} \,. 
\eea
This choice is motivated by the fact that the cosmological constant in the standard model of cosmology must be of background order, in order for it to be influential in the Friedmann equations at late times. There is also no point in perturbing it in either $\epsilon$ or $\eta$, as it is a constant, and the Taylor expansion is trivial. The cosmological constant therefore fits naturally into our two-parameter expansion at lowest-order, as a cosmological background quantity with corresponding scale $L_C^{-2}$.

The full field equations, in arbitrary coordinates, are given in terms of the perturbed quantities introduced in this section in Appendix \ref{explicitfieldequations}. In the sections that follow we will perform gauge transformations in order to determine how these quantities transform between different coordinate systems. We will then construct a set of gauge-invariant quantities that obey the same equations in any coordinate system, before introducing combinations of potentials that can be taken in order to write the gauge-invariant field equations in terms of an effective fluid.

\section{Constructing gauge-invariant variables} \label{Gchoice}

A general infinitesimal gauge transformation between coordinate systems can be written as
\be \label{gt}
x^{\mu} \mapsto \tilde{x}^{\mu} = e^{\xi^{\alpha} \partial_{\alpha}} x^{\mu} \, ,
\ee
where $\xi^{\mu}$ is the gauge generator. All tensors, $\mathcal{T}$, are taken to transform under the gauge transformation in Eq. (\ref{gt}) as
\be \label{gt2}
\tilde{\mathcal{T}} = e^{\mathcal{L}_{\xi}} \mathcal{T} = \mathcal{T} + \mathcal{L}_{\xi} \mathcal{T} + {\textstyle \frac{1}{2}} \mathcal{L}^2_{\xi} \mathcal{T} + \dots \, ,
\ee
where $\mathcal{L}_{\xi}$ denotes the Lie derivative operator with respect to $\xi^{\mu}$. This exponential map results in an invertible transformation, and can be applied to both the metric and the stress-energy tensor. We must now expand the components of the gauge generator in terms of $\epsilon$ and $\eta$, which we do as follows:
\bea
\xi^0 &=& \xi^{(1,0)0} +\xi^{(0,3)0}+\xi^{(1,2)0}+ \dots \sim \epsilon L_C + \eta^3 L_N + \epsilon \eta^3 L_N + \dots \\[10pt]
\xi^i &=& \xi^{(1,0)i} +\xi^{(0,2)i}+\xi^{(1,1)i}+\xi^{(1,2)i}+{\textstyle \frac{1}{2}} \xi^{(0,4)i} \dots \sim \epsilon L_C + \eta^2 L_N + \epsilon \eta^2 L_N + \eta^4 L_N \dots  \, .
\eea
These non-vanishing components of the gauge generator have been chosen so that no new components of the metric or the stress-energy tensor are generated by this transformation, which is an important condition to ensure the problem is being treated in a self-consistent manner.

\subsection{Infinitesimal coordinate transformations}
\label{sec:ict}

In order to perform infinitesimal coordinate transformations it is useful to decompose the perturbed gauge generator into a scalar and a divergence-free vector. Omitting superscripts, these can be written as
\be
\xi^0 \equiv \delta t \qquad {\rm and} \qquad \xi^i \equiv 
{\delta x_,}^{i} + \delta x^i \, ,
\ee
where $\delta x^i_{\phantom{i},i}=0$. The transformation of the metric perturbations due to a gauge transformation of this type are unchanged from the dust-only case, and are given in Section V of Ref. \cite{SRGTCKM}. In the remainder of this section we will outline how the presence of radiation affects the transformation properties of the matter fields $\{ \rho, p ,v_i , \Lambda \}$. This is done using the result from Eq. (\ref{gt2}), and by solving for the decomposed matter variables.

In order to present these results in a form that can be used for cosmology we choose to take $L_N/L_C \sim \eta$. This means that we are restricting the post-Newtonian sector of our expansion to apply on scales below about $100$Mpc, which is realistic also about the size of the homogeneity scale. This is ideal for considering the influence of galaxies, clusters and super-clusters on large-scale linear cosmological perturbations. We also choose, without loss of generality, to express our results in terms of $L_N$. Given this, the total energy density transforms as follows:
\bea
\tilde{\rho}^{(0,0)} +\tilde{\rho}^{(0,2)} &=& \rho^{(0,0)}+ \rho^{(0,2)} \sim \frac{\eta^2}{L_N^2} \label{rho00trans}\\[5pt]
\tilde{\rho}^{(1,1)}&=&\rho^{(1,1)} +\rho^{(0,2)}_{,i}\left(\delta x^{(1,0)\, i}_{\phantom{(1,0)},} + \delta x^{(1,0)i}\right) \label{rho11trans} \sim \frac{\epsilon \eta}{L_N^2} \\[5pt]
\tilde{\rho}^{(1,0)} + \tilde{\rho}^{(1,2)} &=&  \rho^{(1,0)} + \rho^{(1,2)} + \left( \rho^{(0,0)}+ \rho^{(0,2)} \right)^{{\cdot}}\delta t^{(1,0)} \sim \frac{\epsilon \eta^2}{L_N^2} \label{rho1012trans}  \\[5pt]
\tilde{\rho}^{(0,4)} &=&  \rho^{(0,4)} + 2 \rho^{(0,2)}_{,i} \left(\delta x^{(0,2)\, i}_{\phantom{(1,0)},} + \delta x^{(0,2)i}\right) \sim \frac{\eta^4}{L_N^2}  \label{rho04trans}\, , 
\eea
while the total pressure transforms as
\bea
\tilde{p}^{(0,0)} &=& p^{(0,0)} \sim \frac{\eta^2}{L_N^2} \label{p00trans} \\[5pt]
\tilde{p}^{(1,0)} + \tilde{p}^{(1,2)} &=&  p^{(1,0)} + p^{(1,2)}+ \dot{p}^{(0,0)}\delta t^{(1,0)0} -2\frac{\dot{a}}{a}p^{(0,0)}\delta t^{(1,0)0} \sim \frac{\epsilon \eta^2}{L_N^2} \label{p1012trans} \\
\tilde{p}^{(0,4)} &=&  p^{(0,4)} \sim \frac{\eta^4}{L_N^2}\, . \label{p04trans}
\eea
The transformations in Eqs. (\ref{rho11trans}), (\ref{rho04trans}) and (\ref{p04trans}) remain exactly the same as the dust-only case studied in Ref. \cite{SRGTCKM}, while all other transformations are affected by the presence of the radiation. The term $\rho^{(0,0)}$ can be seen to transform in the same was as the Newtonian energy density, $\rho^{(0,2)}$. This is not unexpected, as both quantities have magnitude $\sim L_C^{-2} \sim \eta^2 L_N^{-2}$. Similarly, $\rho^{(0,0)}$ appears alongside $\rho^{(0,2)}$ in the transformation given in Eq. (\ref{rho1012trans}). With the inclusion of radiation, we find that $p^{(0,0)}$ is automatically gauge invariant. Furthermore, as can be seen in Eq. (\ref{p1012trans}), the inclusion of radiation means that the transformation of the cosmological and mixed-order perturbations to the pressure are no longer gauge invariant (as they were in the dust-only case). The reader may note that these results differ from the quasi-static limit of cosmological perturbation theory, as space and time derivatives are treated on a different footing, and because velocities come in at different orders \cite{Peebles}. 

Meanwhile, the peculiar velocities transform in the following way:
\bea
\label{vcostrans}
\tilde{v}_i^{(1,0)} &=& v_i^{(1,0)} - a\left(\delta x^{(1,0)}_{,i}+\delta x^{(1,0)}_i \right)^{\cdot} + v^{(0,1)}_{i,j} \left(\delta x^{(1,0)\, j}_{\phantom{(1,0)},}+\delta x^{(1,0)j} \right) \sim \epsilon \label{v10trans} \\
\tilde{v}_i^{(0,1)} &=& v_i^{(0,1)} \sim \eta \, .\label{v01trans}
\eea
These transformations are the same as in the dust-only case studied in Ref. \cite{SRGTCKM}. Note particularly that in Eq. (\ref{vcostrans}) the small-scale Newtonian velocity contributes to the transformation of the large-scale velocity -- this is a by-product of our two-parameter expansion, and is an effect that would otherwise only appear at second order in standard cosmological perturbation theory. Finally, we find that the cosmological constant $\Lambda^{(0,0)}$ does not transform under the gauge transformation in Eq. (\ref{gt}), as it is a constant in space and time:
\bea
\tilde{\Lambda}^{(0,0)} = \Lambda^{(0,0)} \,. \label{Lambdatrans}
\eea
The transformations above will now be used to construct gauge-invariant quantities.

\subsection{Gauge invariant quantities}
\label{sec:gaugeinvariants}

Let us now create gauge-invariant quantities for the matter degrees of freedom in the presence of radiation and $\Lambda$. Such variables isolate and remove superfluous degrees of freedom, as well as allowing the field equations to be written in a greatly simplified way. To do this it is useful to perform an irreducible decomposition on the metric. Omitting superscripts for simplicity, and without loss of generality, we can do this as follows:
\be
h_{00}  \equiv \phi \, , \qquad h_{0i} \equiv B_{,i} + B_i \qquad {\rm and} \qquad h_{ij} \equiv  - \psi\delta_{ij} + E_{,ij} + F_{(i,j)} + {\textstyle {\frac{1}{2}}} \hat{h}_{ij} \, , \label{metricsplit}
\ee
where $B_i \, , F_i \, , \hat{h}_{ij}$ are divergenceless and $\hat{h}_{ij}$ is trace-free. Applying the gauge transformation (\ref{gt2}) to the metric components (\ref{g00})-(\ref{g0i}) this gives the transformation rules for the irreducibly decomposed components, and allows gauge-invariant gravitational perturbations to be constructed (see Sections V and VI of Ref. \cite{SRGTCKM}). The presence of radiation does not affect the construction of gauge-invariant gravitational perturbations, but does affect the construction of gauge-invariant quantities for the matter variables, which is what we will elaborate upon here.

The method we use to calculate gauge-invariant quantities is as follows: we choose gauge generators $\delta x, \delta x^i $ and $\delta t$ such that the gauge transformed metric potentials $\tilde{E}= \tilde{B} =\tilde{F_i} =0$. We then substitute these gauge generators, now written in terms of $E, B$ and $F_i$, back into the expressions for all of the transformed perturbations presented in Section \ref{Gchoice}. Because the original gauge transformations were written down in a completely arbitrary coordinate system, these new results are automatically gauge invariant \cite{malik}. All such quantities also reduce to metric perturbations in longitudinal gauge when $E=B=F_i = 0$, and have been explicitly checked to be truly gauge invariant.

To construct gauge-invariant matter perturbations we require the transformation laws for $E^{(1,0)}, B^{(1,0)}, F^{(1,0)i}, E^{(0,2)}$ and $F^{(0,2)i}$ under Eq. (\ref{gt}). These are given in \cite{SRGTCKM} and are
\bea
\tilde{B}^{(1,0)} & = & B^{(1,0)} + a \dot{\delta x}^{(1,0)} - {\textstyle \frac{1}{a}} \delta t^{(1, 0)} \sim \epsilon \eta^{-1} L_N  \; \; \; \\
\tilde{E}^{(1, 0)} & = & E^{(1, 0)} + 2 \delta x^{(1, 0)} \sim \epsilon \eta^{-2} L_N^2 \\
\tilde{F}_i^{(1, 0)} & = & F_i^{(1, 0)} + 2 \delta x_i^{(1,0)} \sim \epsilon \eta^{-1} L_N \\
\tilde{E}^{(0,2)} &=& E^{(0,2)} + 2 \delta x^{(0,2)} \sim \eta^2 L_N^2  \\
\tilde{F}_i^{(0,2)} &=& F_i^{(0,2)} + 2 \delta x_i^{(0,2)}\sim \eta^2 L_N \, .   
\eea
For the total energy density perturbations it can then be seen that the following quantities are gauge invariant:
\bea
\mathbf{\rho}^{(0,0)}+ \mathbf{\rho}^{(0,2)} &=&  \rho^{(0,0)} + \rho^{(0,2)} \label{rho02GI}\\[5pt]
\mathbf{\rho}^{(1,1)} &=& \rho^{(1,1)} -\frac{1}{2}\rho^{(0,2)}_{,i}\left(E^{(1,0),i} + F^{(1,0)i}\right) \\[5pt]
\mathbf{\rho}^{(1,0)} + \mathbf{\rho}^{(1,2)} &=&  \rho^{(1,0)} + \rho^{(1,2)}  + \left( \rho^{(0,0)} + \rho^{(0,2)}\right)^{\cdot} \left(aB^{(1,0)} - \frac{a^2}{2}\dot{E}^{(1,0)}\right)  \label{rho1012GI} \\[5pt]
\mathbf{\rho}^{(0,4)} &=&  \rho^{(0,4)} - \rho^{(0,2)}_{,i}\left(E^{(0,2),i} +F^{(0,2)i}\right) \, .
\eea
Correspondingly, for the pressure perturbations we find the following gauge-invariant quantities:
\bea
\mathbf{p}^{(0,0)} &=& p^{(0,0)} \\[5pt]
\mathbf{p}^{(1,0)}+ \mathbf{p}^{(1,2)} &=& p^{(1,0)}+ p^{(1,2)} +\left( \dot{p}^{(0,0)} -2 \frac{\dot{a}}{a}p^{(0,0)} \right)\left( aB^{(1,0)} -\frac{a^2}{2}\dot{E}^{(1,0)} \right)  \\[5pt]
\mathbf{p}^{(0,4)} &=&  p^{(0,4)} \, ,
\eea
and for the peculiar velocity we construct
\bea
\mathbf{v}^{(0,1)}_i &=& v^{(0,1)}_i  \\[5pt]
\mathbf{v}_i^{(1,0)} &=&  
v_i^{(1,0)} + \frac{a}{2}\left(\dot{E}^{(1,0)}_{,i} + \dot{F}_i^{(1,0)} \right)   - \frac{1}{2}v^{(0,1)}_{i,j}\left(E^{(1,0),j} + F^{(1,0)j}\right) \, . \label{V10invar} 
\eea
These last two quantities can be separated into scalar and divergenceless vector parts in a straightforward way. Finally, the gauge-invariant cosmological constant is trivial to construct:
\bea
\mathbf{\Lambda} = \Lambda^{(0,0)} \, . 
\eea
There are no further quantities to consider in the stress-energy tensor, so, when combined with the set of gauge-invariant metric potentials $\{ \Phi, \Psi, \mathbf{B}^{i}, \mathbf{h}_{ij}\}$, constructed in Section VI of Ref. \cite{SRGTCKM}, this gives us a full set of gauge-invariant quantities in our two-parameter perturbative expansion. The field equation in terms of these gauge-invariant variables are given in Appendix \ref{FieldEquationsGaugeInvariantVariables}.

\section{Constructing the field equations}
\label{sec:fe}

The two-parameter expansion described in the previous sections could in principle be applied to numerous different physical systems. While the perturbed metric and stress-energy tensor can be written down without specifying any specific relationship between either $\epsilon$ and $\eta$ or $L_C$ and $L_N$, we must choose how to express these quantities in terms of one another if we want to be able to solve a hierarchical set of field equations. In order to model a realistic universe that has non-linear structure on scales up to $\sim 100$Mpc, as well as linear structure on large scales, we choose $L_N/L_C \sim \eta$. On the other hand, to model a realistic universe, gravitational potentials must have similar magnitude on both small and large scales, so we choose $\epsilon \sim \eta^2$, see Section III of Ref. \cite{SRGTCKM} for justification of this. Both of these requirements are therefore satisfied by the choice
\be
\epsilon \sim \eta^2 \sim \frac{L_N^2}{L_C^2} \sim 10^{-5} \, , \label{choice} 
\ee
where $10^{-5}$ is the typical depth of a potential on both cosmological and post-Newtonian scales. With these relations we can translate our two-parameter expansion into effectively a single-parameter expansion in $\eta$, and write the field equations order-by-order in $\eta$. We further choose to express the field equations in units of $L_N^{-2}$. This last choice has no particular physical significance, and is purely for expediency. We use Eq. (\ref{choice}) in the work that follows, as well in Section \ref{sec:pfe} and Appendices \ref{explicitfieldequations} and \ref{FieldEquationsGaugeInvariantVariables}.

\subsection{Background and Newtonian cosmological equations} \label{SmallAndLarge}

Within the formalism outlined above, the Friedmann-like equations that govern the evolution of the scale factor $a(t)$, and hence the large-scale expansion of the Universe, are {\it not} independent of the perturbations. This can be seen explicitly in Eqs. (\ref{FINAL0002}) and (\ref{FINALij02}) from Appendix  \ref{FieldEquationsGaugeInvariantVariables}, where the Newtonian mass density and gravitational potential act as sources for the cosmological expansion. This is in some sense a very pleasing result; the large-scale expansion of space is driven by the same Newtonian mass that governs the leading-order part of the gravitational field on small scales. On the other hand, it means that our ``background'' is not by itself an exact solution of Einstein's equations. This stretches the meaning of what is usually implied by the phrase ``perturbation theory'' in Einstein's theory\footnote{We are grateful to Marco Bruni for a number of stimulating discussions on this point.}. Nevertheless, both the fundamental objects being perturbed and the field equations themselves are being consistently expanded in the perturbative parameters $\epsilon$ and $\eta$, and we see no reason to expect this expansion should not converge. Indeed the present expansion seems to have much better convergence properties than the standard approach to cosmological perturbation theory, in the presence of non-linear structures \cite{Rasanen}. Furthermore, a change of coordinates on a sub-horizon-sized region of space can be shown to be isometric to perturbed Minkowski space, with the cosmological expansion arising from boundary conditions at the edge of the region \cite{cwoa}. In this sense, the cosmological expansion can be considered an emergent property, and the background on small-scales could equally well be considered to be either a Friedmann model or Minkowski space (which definitely is a solution when $\epsilon=\eta =0$).

In any case, we can now proceed in a similar manner to Section VIIA of Ref. \cite{SRGTCKM} to find the simplest way in which to express the equations that govern the large-scale expansion of space. In order to do this, it us useful to calculate the average mass density and radiation density on distances above the homogeneity scale, $L_{\mathrm{hom}} \sim 100\mathrm{Mpc}$ \cite{Hogg}. These are given by 
\bea
\overline{ \mathbf{\rho}}_M &\equiv & \frac{\int_{V_{\mathrm{hom}}} \mathbf{\rho}^{(0,2)} dV }{\int_{V_{\mathrm{hom}}} dV} \qquad {\rm and} \qquad 
\overline{ \mathbf{\rho}}_R \equiv  \frac{\int_{V_{\mathrm{hom}}} \mathbf{\rho}^{(0,0)} dV }{\int_{V_{\mathrm{hom}}} dV}  =  \, \mathbf{\rho}^{(0,0)}, \label{rhobar2}
\eea
where $V_{\mathrm{hom}}$ indicates the spatial volume associated with the homogeneity scale. Of course, we know from Eq. (\ref{p00timedep}) that there can be no small-scale inhomogeneities in the radiation fluid. For the matter fluid, on the other hand, small-scale fluctuations most definitely do exist and are of order unity. To accommodate these fluctuations we define
\bea
\delta \mathbf{\rho}^{(0,2)}  & \equiv & \mathbf{\rho}^{(0,2)}  -  \overline{\mathbf{\rho}}_M  \, .
\eea
This equation implies that the leading-order inhomogeneous part of the matter energy density, $\delta \mathbf{\rho}^{(0,2)}$, is formally of the same order as the background component of the matter fields, $\mathbf{\rho}^{(0,2)}$, both being $O (\eta^2 L_N^{-2})$. These quantities can now be used to write Eqs. (\ref{FINAL0002}) and (\ref{FINALij02}) into a more useful form.

To derive a set of effective Friedmann equations we first integrate Eq. (\ref{FINALij02}) over the volume corresponding to the homogeneity scale:
\bea \label{intefe}
&& \int_{V_{\textrm{hom}}}\left( 3H^2 -  \frac{1}{a^2}\nabla^2 \Phi^{(0,2)}  \right) dV 
= \int_{V_{\textrm{hom}}} \left( 8 \pi\left(\mathbf{\rho}^{(0,0)} +\mathbf{\rho}^{(0,2)}\right) +\mathbf{\Lambda} \right) dV \, , 
\eea
where $H \equiv \dot{a}/a$. Using Gauss' theorem this can be written as
\bea \label{intefe2}
&&3H^2 V_{\textrm{hom}} -\frac{1}{a^2}\int_{S_{\textrm{hom}}} \nabla \Phi^{(0,2)} \cdot dS 
= 8 \pi \left(\overline{ \mathbf{\rho}}_M + \overline{ \mathbf{\rho}}_R \right) V_{\text{hom}}+ \mathbf{\Lambda} V_{\text{hom}} \, . 
\eea
If we now assume that on the homogeneity scale there is no net flux of $\nabla \Phi^{(0,2)}$ into or out of the surface $S_{\textrm{hom}}$, then the second term in Eq. (\ref{intefe2}) vanishes. This leaves us with
\begin{equation}
H^2 = \frac{8 \pi}{3} \left( \overline{ \mathbf{\rho}}_M + \overline{ \mathbf{\rho}}_R \right) + \frac{\mathbf{\Lambda}}{3} \, , \label{homolowest}
\end{equation}
which is exactly the same form as the standard Friedmann equation in the presence of matter, radiation and a cosmological constant. What is more, the lowest-order parts of the stress-energy conservation equations yields the results \cite{sanghai2}
\be
\overline{ \mathbf{\rho}}_M \propto a^{-3} \qquad {\rm and} \qquad \overline{ \mathbf{\rho}}_R \propto a^{-4} \, ,
\ee
which are again exactly as expected from Friedmann cosmology. Substituting these results back into Eq. (\ref{FINALij02}) gives
\begin{equation} 
\nabla^2 \Phi^{(0,2)} = - 8 \pi a^2 \delta \mathbf{\rho}^{(0,2)} \, , \label{inhomolowest}
\end{equation}
which is identical to the standard equation used in Newtonian N-body simulations for cosmology. In summary, we find that the leading-order parts of the field equations, in the context of our two-parameter expansion, reproduce exactly the same results as standard Friedmann cosmology with dust, radiation and a cosmological constant (although the meaning of the equations is slightly different). In the following section we will find that this is not the case when the non-linear aspects of Einstein's equations become important, on large scales.

Finally, we note that the Friedmann equation is recovered from our expansion when $\delta \mathbf{\rho}^{(0,2)} =0$ (i.e. when the leading-order contribution to the energy density is homogeneous). This is not the same condition as setting $\epsilon=\eta=0$, which would correspond to an empty space within our framework.

\subsection{Leading-order cosmological perturbation equations}

The equations presented in Appendix \ref{FieldEquationsGaugeInvariantVariables} constitute a hierarchy of field equations, where the equations from Section \ref{SmallAndLarge} are the leading-order parts. Once the Friedmann equation (\ref{homolowest}) and the Newtonian equation (\ref{inhomolowest}) have been solved, then their solutions can be substituted into the remaining higher-order equations to gain a set of solutions for the leading-order cosmological perturbations. This latter set of solutions, at $\mathcal{O}(\eta^4 L_N^{-2})$, contain linear-order cosmological large-scale potentials and post-Newtonian potentials from small scales. With this in mind, we therefore seek to recast the $\mathcal{O}(\eta^4 L_N^{-2})$ equations in the form of the equations of standard first-order cosmological perturbation theory, modified by the addition of terms related to the existence of inhomogeneity on the length scale $L_N$. These terms will be then be treated as forming the components of an \textit{effective fluid} on large scales, whose characteristics and behaviour is determined by the small-scale gravitational physics. Such an approach has similarities to the effective fluid approaches in for example \cite{Baumann:2010tm, Senatore}, but in our case is also required to reduce the number of gravitational degrees of freedom to be no more than the available number of field equations.

In the end, we want to reduce to a set of six perturbed field equations for six degrees of freedom (i.e. the 10 degrees of freedom in the metric minus the four coordinate freedoms). At present, Eqs. (\ref{FINAL0i03})-(\ref{FINALijtracefree04}) from Appendix \ref{FieldEquationsGaugeInvariantVariables} contain a total of sixteen degrees of freedom: six scalars ($\Phi^{(1,0)}$, $\Phi^{(1,2)}$, $\Phi^{(0,4)}$, $\Psi^{(1,0)}$, $\Psi^{(1,2)}$ and $\Psi^{(0,4)}$), six in the tensors ($\textbf{h}_{ij}^{(1,0)}$, $\textbf{h}_{ij}^{(1,2)}$ and $\textbf{h}_{ij}^{(0,4)}$) and four in the vectors ($\textbf{B}^{(1,0)}$ and $\textbf{B}^{(1,2)}$). Taking into account the four degrees of freedom removed by gauge fixing implies that we need to remove six degrees of freedom. This is achieved by defining new sets of variables as follows:
\bea
\label{new1}
U &\equiv& - {\textstyle \frac{1}{2}} \left( \Phi^{(0,2)} + \Phi^{(1,1)} \right)\\
\upphi &\equiv& - {\textstyle \frac{1}{2}} \left( \Phi^{(1,0)} + \Phi^{(1,2)} +{\textstyle \frac{1}{2}}\Phi^{(0,4)} \right) \\
\uppsi &\equiv& {\textstyle \frac{1}{2}} \left(\Psi^{(1,0)} + \Psi^{(1,2)} + {\textstyle \frac{1}{2}}\Psi^{(0,4)} \right) \\
 \mathrm{S}_j &\equiv& -\left( \textbf{B}_j^{(1,0)} +\textbf{B}_j^{(0,3)}+ \textbf{B}_j^{(1,2)} \right) \\
{\rm h}_{ij} &\equiv& {\textstyle \frac{1}{4}} \left( \textbf{h}_{ij}^{(1,0)} + \textbf{h}_{ij}^{(1,2)} +{\textstyle \frac{1}{2}}\textbf{h}_{ij}^{(0,4)} \right) \, , 
\eea
and
\bea
\updelta \uprho_{\rm N} &\equiv& \delta \mathbf{\rho}^{(0,2)} + \mathbf{\rho}^{(1,1)} \\
\updelta \uprho &\equiv& \mathbf{\rho}^{(1,0)} +\mathbf{\rho}^{(1,2)} +{\textstyle \frac{1}{2}}\mathbf{\rho}^{(0,4)} \\
\updelta {\rm p} &\equiv& \mathbf{p}^{(1,0)} + \mathbf{p}^{(1,2)} + {\textstyle \frac{1}{2}} \mathbf{p}^{(0,4)}\\
{\rm v}_{{\rm N}i} &\equiv& \mathbf{v}^{(0,1)}_i  \\
{\rm v}_{i} &\equiv& \mathbf{v}^{(1,0)}_i \, ,
\label{new10}
\eea
which are exactly the sets of variables used in Section \ref{sec:pfe}. A number of these new variables could be considered to be ``composite quantities'', as they contain a number of different perturbative orders in the same variable. For example, the variable $\uppsi$ is dominated by $\mathcal{O}(\epsilon)$ terms on cosmological length scales $L_C$, but contains smaller terms at $\mathcal{O}(\eta^4)$ on small-scales $L_N$. This is quite atypical in perturbation theory. However, the way in which these quantities arise together in the field equations suggest that they should be solved for together.

The equations displayed in Section \ref{sec:pfe} can now be seen to be given by those from Appendix \ref{FieldEquationsGaugeInvariantVariables}, but written in terms of the variables in Eqs. (\ref{new1})-(\ref{new10}) and in terms of conformal time $d \tau \equiv a^{-1} dt$. To solve this system, one must solve the Newton-Poisson equation and Friedmann equations for $U$ and $a(t)$, respectively. This should not be too challenging, as they are identical to the standard equations used in Newtonian N-body simulations for Friedmann cosmology. From these results one can then solve for the effective fluid parameters, and then solve the cosmological perturbation equations. This will inevitably be complicated by the additional ``mode-mixing'' terms in the cosmological perturbation equations, which will require more sophisticated techniques than at leading-order in standard cosmological perturbation theory. These terms suggest that it may in fact be possible to generate vector and tensor modes from scalar fluctuations, which is already well known in second-order cosmological perturbation theory \cite{Malik:2003mv, Nakamura:2006rk}, but is not usually seen at first order. One should also note that these terms, for example $\frac{8 \pi a^2}{3} \updelta \uprho_{\rm N} (\uppsi - \upphi)$, also mean that Fourier modes no longer decouple in a trivial way as they do in standard first-order perturbation theory, even if no mode-mixing occurs. This is because the Fourier transforms of such terms are expressible only in terms of a convolution integral over all Fourier modes.

\section{Conclusions} \label{conc}

We construct a two-parameter perturbation expansion around an FLRW background that simultaneously describes non-linear structures on small-scales and linear structures on large scales. Moreover, it includes matter, radiation and a cosmological constant. In doing so we use both cosmological perturbation theory and the post-Newtonian expansion. As this expansion is able to model large density contrasts and different matter components it therefore both contains the essential features of the real Universe and has a number of potential advantages over standard cosmological perturbation theory.
We derived the two-parameter perturbed field equations valid for structure on the order of a fraction on the horizon size, the two-parameter gauge transformations of the matter sector, and construct gauge-invariant quantities in this sector. The consistency of the gauge transformations requires not only gravitational potentials and matter perturbations at the orders expected from post-Newtonian gravity and cosmological perturbation theory alone, but also a number of others at orders in perturbation which may not naively have been expected. We have therefore identified a minimal set of perturbations that are required for mathematical consistency of the problem, and written down gauge-invariant versions of the field equations that contain them all. 

We find that the small-scale Newton-Poisson equation for the scalar gravitational potential occurs at the same order in perturbations as the Friedmann equation, but that they can be separated after the introduction of a suitable homogeneity scale. At leading order, this results in a small-scale Newton-Poisson equation sourced by the inhomogeneous part of the Newtonian energy density, and large-scale Friedmann equations sourced by the spatial average of the leading-order parts of the energy density, pressure, and the cosmological constant. Our results give no indications that the effects of small-scale non-linearities should be expected to cause acceleration of the large-scale Universe, but we do find that they should be expected to affect large-scale perturbations. This is because the higher-order field equations include quadratic Newtonian potentials within the effective fluid terms. They therefore contain valuable information about non-linear gravity, and could potentially be used to identify relativistic effects in observations of large-scale structure.

By presenting the higher-order field equations in terms of an effective fluid we are able to highlight the similarities and differences between our formalism and regular cosmological perturbation theory. We expect this to aid further application of our equations by allowing some standard techniques from cosmological perturbation theory to be imported. Our effective fluid description also enables an easier physical interpretation of the effects of non-linearities in the field equations, which clearly lead to (for example) a large-scale effective pressure and anisotropic stress. Since the effective fluid terms are all constructed from the solution to the short-scale Newtonian gravitational potential, their properties should be able to be determined from N-body simulations. Once the form of these effective fluids has been identified, one can proceed to solve the cosmological equations for the long-wavelength perturbations. This method of solution is available to us because of the hierarchical nature of the perturbation equations -- short-scale fluctuations appear at lower-order compared to cosmological perturbations, and so can be solved for before cosmological perturbations. Within this prescription we observe a mixing of scales, as well as mode-mixing at what would normally be considered to be linear order in cosmological potentials. Understanding the consequences of these relativistic effects on the formation of non-linear structure in the Universe is of importance not only for removing sources of observational bias, but also because it has the potential to offer new ways of probing Einstein's theory with cosmology. 

\section*{Acknowledgements}

\noindent
We are grateful to Karim Malik and Marco Bruni for helpful discussions. SRG , CSG and TC are supported by the STFC under grants ST/K50225X/1 and ST/N504257/1. The tensor algebra packages xAct \cite{xAct} and xPand \cite{xPand1, xPand2} were used in this work. 

\appendix
\section{The two-parameter perturbed stress-energy tensor} \label{appendixlong}

This appendix provides expressions for the perturbed stress-energy tensor, used to derive the field equations in Sections \ref{sec:pfe} and Appendices \ref{explicitfieldequations} and \ref{FieldEquationsGaugeInvariantVariables}. The perturbed Ricci tensor is unchanged from the dust-only case, and is given in the Appendix of Ref. \cite{SRGTCKM}. The equations below make no assumptions about the relative magnitude of $\epsilon$ and $\eta$, nor is anything assumed about the length scales $L_C$ and $L_N$. Expanding the total stress-energy tensor in both $\epsilon$ and $\eta$ gives the non-vanishing components as follows:
\bea 
T_{00} &=& T^{(0,0)}_{00} + T^{(0,2)}_{00}  + {\textstyle \frac{1}{2}} T^{(0,4)}_{00}  + T_{00}^{(1,0)} + T^{(1,1)}_{00}+ T^{(1,2)}_{00} + \ldots \qquad \quad \label{T00pert}  \\[5pt] 
T_{ij} &=&  T^{(0,0)}_{ij}+T^{(0,2)}_{ij}   + T^{(1,0)}_{ij}+ T^{(1,1)}_{ij}+ T^{(1,2)}_{ij} + {\textstyle \frac{1}{2}} T^{(0,4)}_{ij} \ldots \label{Tijpert} \\[5pt]
T_{0i} &=&  T_{0i}^{(0,1)}+ T_{0i}^{(0,3)} + T_{0i}^{(1,0)} + T_{0i}^{(1,2)} + \ldots \label{T0ipert} \, .
\eea
The terms on the right-hand side of Eq. (\ref{T00pert}) are given by
\bea
T^{(0,0)}_{00} &=&  \rho^{(0,0)} \sim \frac{1}{L_C^2}  \label{T0000} \\
T^{(0,2)}_{00} &=&  \rho^{(0,2)} -\rho^{(0,0)}h^{(0,2)}_{00} +(\rho^{(0,0)} +p^{(0,0)})v^{(0,1)i}v^{(0,1)}_i 
\sim \frac{\eta^2}{L_N^2} + \frac{\eta^2}{L_C^2} \label{T0002} \\
T^{(0,4)}_{00} &=& \frac{1}{2} \rho^{(0,4)} - h^{(0,2)}_{00}\rho^{(0,2)} \label{T0004}  +  \rho^{(0,2)} v^{(0,1)i}v^{(0,1)}_i  + \, \mathrm{terms \, of \, size \, } \left[ \frac{\eta^4}{L_C^2} \right] \sim  \frac{\eta^4}{L_N^2} + \frac{\eta^4}{L_C^2}  \\
T^{(1,0)}_{00} &=&  \rho^{(1,0)} - \rho^{(0,0)}h^{(1,0)}_{00} \sim \frac{\epsilon}{L_C^2} \label{T0010} \\
T^{(1,1)}_{00} &=&  \rho^{(1,1)} + \, \mathrm{terms \, of \, size \, } \left[ \frac{\epsilon \eta}{L_C^2} \right] \label{T0011} 
\sim  \frac{\epsilon \eta}{L_N^2} + \frac{\epsilon \eta}{L_C^2}  \\
T^{(1,2)}_{00} &=&  \rho^{(1,2)} - h^{(1,0)}_{00}\rho^{(0,2)} + \, \mathrm{terms \, of \, size \, } \left[ \frac{\epsilon \eta^2}{L_C^2} \right] \label{T0012}  
\sim  \frac{\epsilon \eta^2}{L_N^2} +\frac{\epsilon \eta^2}{L_C^2}  \, , 
\eea
while the terms in Eq. (\ref{Tijpert}) are given by
\bea
T^{(0,0)}_{ij} &=&   a^2 p^{(0,0)} \delta_{ij}  \sim \frac{1}{L_C^2}  \label{Tij00} \\
T^{(0,2)}_{ij} &=&   a^2 \left(\rho^{(0,0)} + p^{(0,0)} \right)v^{(0,1)}_i v^{(0,1)}_j  + a^2 p^{(0,0)}h^{(0,2)}_{ij}  \sim \frac{\eta^2}{L_C^2}\,  \label{Tij02} \\
T^{(1,0)}_{ij} &=&   a^2 p^{(1,0)}\delta_{ij} +a^2 h^{(1,0)}_{ij} p^{(0,0)} \sim \frac{\epsilon}{L_C^2}  \label{Tij10} \\
T^{(1,1)}_{ij} &=&   \mathrm{terms \, of \, size \, }\left[ \frac{\epsilon \eta}{L_C^2}\right]  \label{Tij11} \\
T^{(1,2)}_{ij} &=&   a^2 p^{(1,2)}\delta_{ij} + \mathrm{terms \, of \, size \, }\left[ \frac{\epsilon \eta^2}{L_C^2}\right]  \label{Tij10} \sim   \frac{\epsilon \eta^2}{L_N^2} + \frac{\epsilon \eta^2}{L_C^2}  \\
T^{(0,4)}_{ij} &=&   a^2 \rho^{(0,2)}v^{(0,1)}_i v^{(0,1)}_j + \frac{a^2}{2} p^{(0,4)} \delta_{ij}+  \, \mathrm{terms \, of \, size \, } \left[ \frac{\eta^4}{L_C^2} \right] \label{Tij04} 
\sim \frac{\eta^4}{L_N^2} +\frac{\eta^4}{L_C^2}\, ,
\eea
and the terms in Eq. (\ref{T0ipert}) are given by
\bea
T^{(0,1)}_{0i} &=&  -a (\rho^{(0,0)} + p^{(0,0)})v_i^{(0,1)} \sim \frac{\eta}{L_C^2} \label{T0i01} \\
T^{(0,3)}_{0i} &=&  -a \rho^{(0,2)}v_i^{(0,1)} + \, \mathrm{terms \, of \, size \, } \left[ \frac{\eta^3}{L_C^2} \right]  \label{T0i03}  
\sim \frac{\eta^3}{L_N^2} +\frac{\eta^3}{L_C^2}  \\
T^{(1,0)}_{0i} &=&  -a \rho^{(0,0)}(v_i^{(1,0)} + h^{(1,0)}_{0i}) -a p^{(0,0)}v^{(1,0)}_i
\sim  \frac{\epsilon}{L_C^2} \label{T0i10} \\
T^{(1,2)}_{0i} &=&  -a\rho^{(0,2)}\left(v^{(1,0)}_i + h^{(1,0)}_{0i}\right)  \label{T0i12} 
-a \rho^{(1,1)}v^{(0,1)}_i + \, \mathrm{terms \, of \, size \, } \left[ \frac{\epsilon \eta^2}{L_C^2} \right]
\sim \frac{\epsilon \eta^2}{L_N^2} + \frac{\epsilon \eta^2}{L_C^2}\, .
\eea
This completes the list of expanded stress-energy tensor components required for Sections \ref{sec:pfe} and Appendices \ref{explicitfieldequations} \& \ref{FieldEquationsGaugeInvariantVariables}.

\section{Field equations without gauge-fixing} \label{explicitfieldequations}

This appendix contains the field equations in terms of the variables introduced in Section \ref{sec:ThFr}, with the choice of relations between $\epsilon$, $\eta$, $L_C$ and $L_N$ given in Eq. (\ref{choice}).

\subsection{Background-order potentials} \label{bkefes}

\noindent
The leading-order $00$-field equation is of order $\mathcal{O}(\eta^2 L_{N}^{-2})$ and is given by 
\begin{equation} \label{e00N}
3\frac{\ddot{a}}{a} + \frac{1}{2 a^2} \nabla^2 h^{(0,2)}_{00} = - 4 \pi  \left(\rho^{(0,0)} +\rho^{(0,2)} + 3p^{(0,0)} \right) +  \Lambda^{(0,0)} \,.
\end{equation}
Note that $a \sim 1$ and $\ddot{a} \sim 1/L_C^2$, as the time variation of $a(t)$ is over cosmological scales. This equation is a combination of the Raychaudhuri equation from Friedmann cosmology, and the Newton-Poisson equation from post-Newtonian gravity. The leading-order contribution to the trace of the $ij$-field equation  at $\mathcal{O}(\eta^2 L_{N}^{-2})$ is given by
\bea
\left(\frac{\dot{a}}{a}\right)^2 - \frac{1}{6 a^2} \left(\nabla^2 h^{(0,2)}_{ii} -h^{(0,2)}_{ij,ij} \right) = \frac{8\pi}{3} \left(\rho^{(0,2)} +\rho^{(0,0)} \right) +\frac{1}{3}\Lambda^{(0,0)} \, . \label{ijtrace}
\eea
This equation is a combination of the Friedmann constraint equation and the Newton-Poisson equation for the trace of the post-Newtonian potential $h_{ii}^{(0,2)}$. At the same order, the trace-free part of the $ij$-field equations is
\bea
D_{ij} \left( h^{(0,2)}_{00} - h^{(0,2)}_{kk} \right) + 2 h^{(0,2)}_{k\langle i,j\rangle k} - \nabla^2 h^{(0,2)}_{\langle ij \rangle} = 0 \, . \label{ijtracefree}
\eea
This equation is the same as given in Ref. \cite{SRGTCKM} because neither the cosmological constant nor radiation contribute trace-free components. The other equations contain contributions from both radiation and the cosmological constant.

\subsection{Vector potentials} \label{bbkefes}

\noindent
The $0i$-field equations give the governing equations for the vector gravitational potentials. The leading-order contribution is $\mathcal{O}(\eta^3 L_N^{-2})$ and is given by
\bea \label{e0iPN}
\nabla^2 h^{(0,3)}_{0i} -  h^{(0,3)}_{0j,ij} - a \dot{h}^{(0,2)}_{ij,j} + a \dot{h}^{(0,2)}_{jj,i} + 2\dot{a} h^{(0,2)}_{00,i} = 16 \pi a^2 \left(\rho^{(0,0)} + \rho^{(0,2)} + p^{(0,0)}\right) v^{(0,1)}_i \,.
\eea
This is the equation for the small-scale post-Newtonian vector potential, responsible for phenomena such as the Lense-Thirring effect, and is the one studied in Ref. \cite{bruni}. Interestingly, this field equation implies the gravitomagnetic potential is $\sim 100$ times larger than second-order perturbation theory predicts \cite{SRGTCKM, AndEtAl}. The next-to-leading-order $0i$-field equation occurs at $\mathcal{O}(\eta^4 L_N^{-2})$, and is given by 
\bea \label{0i1.5efe}
\hspace{-4pt}
&& \nabla^2 \left(h^{(1,0)}_{0i} +h^{(1,2)}_{0i}\right)- \left(h^{(1,0)}_{0j} + h^{(1,2)}_{0j}\right)_{,ij}  -  h^{(1,0)}_{0j}h^{(0,2)}_{00,ij}   - a \left(h^{(1,0)}_{ij}+h^{(1,1)}_{ij}\right)^{\cdot}_{,j} 
+ a  \left(h^{(1,0)}_{jj}+h^{(1,1)}_{jj}\right)^{\cdot}_{,i} \\
&&+2 \dot{a} \left(h^{(1,0)}_{00}+h^{(1,1)}_{00}\right)_{,i}  - 2 h^{(1,0)}_{0i} \left( 2 \dot{a}^2  + a \ddot{a} \right) \nonumber \\
&=&
8 \pi a^2  \left(2\left( \rho^{(0,0)} + \rho^{(0,2)} +p^{(0,0)}  \right)v^{(1,0)}_i +  2\rho^{(1,1)}v^{(0,1)}_i + \left(\rho^{(0,0)} +\rho^{(0,2)} + 3p^{(0,0)}\right) h^{(1,0)}_{0i} \right) -2a^2 \Lambda^{(0,0)}h_{0i}^{(1,0)} \,. \nonumber
\eea
This equation is the governing equation for the large-scale vector potentials. It is more complicated than Eq. (\ref{e0iPN}), and shows that non-linear gravitational effects could potentially source the growth of large-scale vector potentials at late times. This equation can also be seen to have contributions from radiation and the cosmological constants, unlike Eq. (\ref{e0iPN}).

\subsection{Higher-order scalar potentials}

\noindent
The next-to-leading-order $00$-field equation occurs at $\mathcal{O}( \eta^3 L_N^{-2})$, and given by
\begin{equation} \label{e0011}
\nabla^2 h^{(1,1)}_{00} = - 8 \pi a^2 \rho^{(1,1)} \, .
\end{equation}
This is another version of the Newton-Poisson equation, and is sourced only by a mixed-order matter energy density, $\rho^{(1,1)}$. The governing equations for the cosmological potentials $h^{(1,0)}_{00}$ and $h^{(1,0)}_{ii}$ occur along with post-Newtonian and mixed-order potentials at $\mathcal{O}(\eta^4 L_N^{-2})$ (as was the case for the vector potentials considered above). The $00$-field equation at this order gives
\bea \label{e00PN}
&&\nabla^2 \left( h^{(1,0)}_{00} + h^{(1,2)}_{00} + \frac{1}{2}h^{(0,4)}_{00} \right) 
+ \frac{1}{2} \left(\nabla h^{(0,2)}_{00} \right)^2  
+ a^2 \left( h^{(0,2)}_{ii} + h^{(1,0)}_{ii} \right)^{\cdot \cdot}
- 2 \left[ a \left( h^{(0,3)}_{0i}  + h^{(1,0)}_{0i} \right)_{,i} \right]^{\cdot}  
\nonumber \\[5pt] 
&& + 2a \dot{a} \left( h^{(0,2)}_{ii} +h^{(1,0)}_{ii} \right)^{\cdot} 
- \frac{1}{2} h^{(0,2)}_{00,i} \left( 2h^{(0,2)}_{ij,j} -h^{(0,2)}_{jj,i} \right) 
- h^{(0,2)}_{00,ij} \left( h^{(1,0)}_{ij} + h^{(0,2)}_{ij} \right) 
+ 3 a \dot{a} \left( h^{(0,2)}_{00} +h^{(1,0)}_{00} \right)^{\cdot} 
\nonumber \\[5pt] 
&=& -8 \pi a^2  \left[ \rho^{(1,0)}  + \rho^{(1,2)} + \frac{1}{2}\rho^{(0,4)} - \left(\rho^{(0,0)}+ \rho^{(0,2)} +3p^{(0,0)} \right) \left( h^{(1,0)}_{00} + h^{(0,2)}_{00} \right)  + 3 \left( p^{(1,0)} + p^{(1,2)} +\frac{1}{2} p^{(0,4)} \right) \right] \nonumber \\
&& -16\pi a^2 \left( v^{(0,1)}_i \right)^2 \left(\rho^{(0,0)} + \rho^{(0,2)} +p^{(0,0)} \right) - 2a^2\Lambda^{(0,0)}\left( h^{(0,2)}_{00} +h^{(1,0)}_{00} \right)  \, . 
\eea 
This equation can be seen to have additional sources due to the presence of radiation and a cosmological constant, compared to the corresponding equation in the presence of dust only \cite{SRGTCKM}. The next non-trivial order in the $ij$-field equation is at $\mathcal{O}(\eta^3 L_N^{-2})$. Its trace gives
\bea
\nabla^2 h^{(1,1)}_{ii} -h^{(1,1)}_{ij,ij}  = - 16\pi a^2 \rho^{(1,1)} \, , \label{ij11trace}
\eea
and its trace-free part is given below. Similarly the $ij$-field equation at $\mathcal{O}(\eta^4 L_N^{-2})$ can also be split into its trace and trace-free parts. The trace of this equation gives 
\bea 
&& \left(\delta_{ij} \nabla^2- \partial_i \partial_j\right) \left( h^{(1,0)}_{ij} + h^{(1,2)}_{ij} + \frac{1}{2}h^{(0,4)}_{ij} \right) 
- \left(2\dot{a}^2 +   a \ddot{a}\right) \left( h_{ii}^{(1,0)} +h_{ii}^{(0,2)}+ 3 h^{(1,0)}_{00}+ 3 h^{(0,2)}_{00} \right)  \nonumber \\[5pt] 
&& +4 \dot{a}\left( h^{(1,0)}_{0i} + h^{(0,3)}_{0i} \right)_{,i}  
 -2a\dot{a}\left( h_{ii}^{(1,0)}+ h_{ii}^{(0,2)} \right)\dot{} \nonumber \\[5pt] 
&=& -16 \pi a^2 \left[   \rho^{(1,0)} +  \frac{1}{2}\rho^{(0,4)} + \rho^{(1,2)} +\left(\rho^{(0,0)} +\rho^{(0,2)} +p^{(0,0)}\right)\left( v^{(0,1)}_i \right)^2  \right] \nonumber \\[5pt]
&& -4 \pi a^2 \left[ \left( \rho^{(0,0)} + \rho^{(0,2)} -p^{(0,0)} \right)\left(h^{(0,2)}_{ii} + h^{(1,0)}_{ii} \right) - \left( \rho^{(0,0)} +\rho^{(0,2)} +3p^{(0,0)} \right)\left( h^{(0,2)}_{00} +h^{(1,0)}_{00} \right)  \right] \nonumber \\[5pt]
&& -a^2\Lambda^{(0,0)}\left[h^{(0,2)}_{00}+h^{(1,0)}_{00}+ h^{(0,2)}_{ii}+ h^{(1,0)}_{ii} \right]+ \mathcal{A} \, , \label{eijPNtrace}
\eea
where we have simplified using Eq. (\ref{e00PN}), and where $\mathcal{A}$ is given by
\bea 
\mathcal{A} 
&\equiv & 
\frac{3}{4} \left( h^{(0,2)}_{ij, k} \right)^2 
+ h^{(0,2)}_{ij, j} \left( h^{(0,2)}_{kk,i} -  h^{(0,2)}_{ik,k} \right) 
- \frac{1}{2}h^{(0,2)}_{ij,k} h^{(0,2)}_{ik,j} 
- \frac{1}{4}h^{(0,2)}_{ii,j}h^{(0,2)}_{kk,j} 
+ \frac{1}{2}\nabla^2 h^{(0,2)}_{00} \left( h^{(1,0)}_{00} + h^{(0,2)}_{00} \right) \nonumber \\[5pt] 
&&+ \frac{1}{2} \left( h^{(0,2)}_{00,ij}+ \nabla^2 h^{(0,2)}_{ij} \right) \left( h^{(1,0)}_{ij} + h^{(0,2)}_{ij} \right) 
+ \left( \frac{1}{2}h^{(0,2)}_{ii,jk} -  h^{(0,2)}_{ij,ik} \right) \left( h^{(0,2)}_{jk} + h^{(1,0)}_{jk} \right)  \, .
\eea
This equation includes new source terms due to the radiation and cosmological constant. The trace-free part of this equation is presented below. 

\subsection{Tensor potentials}

\noindent
The next-to-leading-order trace-free $ij$-field equation occurs at $\mathcal{O}(\eta^3L_N^{-2})$, and is given by
\bea
D_{ij} \left( h^{(1,1)}_{00} - h^{(1,1)}_{kk} \right) + 2 h^{(1,1)}_{k\langle i,j\rangle k} - \nabla^2 h^{(1,1)}_{\langle ij \rangle} = 0 \, . \label{ij11tracefree}
\eea
Finally, the trace-free part of the $ij$-field equation at $\mathcal{O}(\eta^4 L_N^{-2})$ is given by
\bea  \label{eijPNtracefree}
&&
\nabla^2 \left( h^{(1,0)}_{\langle ij \rangle} +h^{(1,2)}_{\langle ij \rangle} + \frac{1}{2}h^{(0,4)}_{\langle ij \rangle} \right) 
- D_{ij} \left( h^{(1,0)}_{00} + h^{(1,2)}_{00} +\frac{1}{2}h^{(0,4)}_{00} - h^{(1,0)}_{kk} - h^{(1,2)}_{kk} - \frac{1}{2}h^{(0,4)}_{kk} \right) 
\nonumber \\[5pt] 
&&
- 2 \left( h^{(1,0)}_{k\langle i} +h^{(1,2)}_{k\langle i} +\frac{1}{2}h^{(0,4)}_{k\langle i} \right)_{,j \rangle k}   
- a^2 \left( h^{(1,0)}_{\langle ij \rangle} + h^{(0,2)}_{\langle ij \rangle } \right)^{\cdot \cdot}
\nonumber \\[5pt] 
&&
- 2 \left(2\dot{a}^2+ a\ddot{a}\right) \left( h^{(1,0)}_{\langle ij \rangle } + h^{(0,2)}_{\langle ij \rangle } \right)  
- 3a\dot{a} \left( h^{(1,0)}_{\langle ij \rangle}   + h^{(0,2)}_{\langle ij \rangle } \right)^{\cdot}   
+ \frac{2}{a} \left[ a^2 \left( h^{(1,0)}_{0\langle i} +h^{(0,3)}_{0\langle i} \right)\right]^{\cdot}_{,j \rangle }  
 \nonumber \\[5pt] 
&=& -8 \pi a^2 \left[ \left( \rho^{(0,0)} +\rho^{(0,2)} -p^{(0,0)} \right)  \left( h^{(0,2)}_{\langle ij \rangle} +h^{(1,0)}_{\langle ij \rangle} \right)+ 2 \left( \rho^{(0,0)} +\rho^{(0,2)} +p^{(0,0)} \right) v^{(0,1)}_{\langle i} v^{(0,1)}_{j\rangle} \right] \nonumber \\[5pt]
&&  -2a^2 \Lambda^{(0,0)}\left( h^{(0,2)}_{\langle ij \rangle} +h^{(1,0)}_{\langle ij \rangle}  \right) + \mathcal{B}_{ij}  \, , 
\eea
where
\bea \label{Bij}
\mathcal{B}_{ij} &\equiv &  
\frac{1}{2} h^{(0,2)}_{00, \langle i \vert}h^{(0,2)}_{00, \vert j \rangle}  
+\frac{1}{2} h^{(0,2)}_{kl, \langle i\vert} h^{(0,2)}_{kl, \vert j \rangle}  
+  D_{ij}h^{(0,2)}_{00} \left( h^{(1,0)}_{00} + h^{(0,2)}_{00} \right)  
+\frac{1}{2} \left( h^{(0,2)}_{00,k} + 2h^{(0,2)}_{kl,l} -h^{(0,2)}_{ll,k} \right) \left( h^{(0,2)}_{\langle ij \rangle,k} -2h^{(0,2)}_{k\langle i,j\rangle} \right) 
\nonumber \\[5pt] 
&&+ \left( D_{ij}h^{(0,2)}_{kl} + h^{(0,2)}_{\langle ij \rangle,kl} -2h^{(0,2)}_{k\langle i,j \rangle l} \right) \left( h^{(1,0)}_{kl} + h^{(0,2)}_{kl} \right)  
+ h^{(0,2)}_{\langle i \vert k,l} \left( h^{(0,2)}_{\vert j\rangle k,l} -h^{(0,2)}_{\vert j\rangle l,k} \right)  \, . 
\eea
This completes the full set of field equations, to the order at which we require them.

\section{Field equations in gauge-invariant variables} \label{FieldEquationsGaugeInvariantVariables}

This appendix contains the field equations in terms of the gauge-invariant variables from Section \ref{sec:gaugeinvariants} and \cite{SRGTCKM}. The choice of relations between $\epsilon$, $\eta$, $L_C$ and $L_N$ is again the same as those given in Eq. (\ref{choice}).

\subsection{Background-order potentials}

\noindent
The trace-free part of the $ij$-equations at $\mathcal{O}(\eta^2L_N^{-2})$ gives
\bea
D_{ij}\left(\Phi^{(0,2)} + \Psi^{(0,2)} \right) - \frac{1}{2}\nabla^2 \mathbf{h}_{ij}^{(0,2)} =0 \, , \label{FINALijnottrace02}
\eea
which implies 
\be
\Phi^{(0,2)} = - \Psi^{(0,2)} \qquad {\rm and} \qquad \mathbf{h}^{(0,2)}_{ij}= 0 \, . \label{psiphihij02}
\ee
The $00$-field equation at $\mathcal{O}(\eta^2L_N^{-2})$ can be written as
\bea
&\ & \frac{\ddot{a}}{a} + \frac{1}{6a^2}\nabla^2 \Phi^{(0,2)} = - \frac{4 \pi}{3}  \left( {\mathbf \rho}^{(0,0)} + {\mathbf \rho}^{(0,2)} +3\mathbf{p}^{(0,0)} \right) +\frac{1}{3}{\mathbf \Lambda} \, , \label{FINAL0002} 
\eea
and the trace of the $ij$-equation at $\mathcal{O}(\eta^2L_N^{-2})$ gives
\bea
&\ & \left( \frac{\dot{a}}{a} \right)^2 - \frac{1}{3 a^2} \nabla^2 \Phi^{(0,2)} = \frac{8 \pi}{3} \left( {\mathbf \rho}^{(0,0)} +{\mathbf \rho}^{(0,2)} \right) +\frac{1}{3}\mathbf{\Lambda} \, ,
\label{FINALij02}
\eea
where we have substituted in the results from Eq. (\ref{psiphihij02}). These equations govern the leading-order part of the gravitational field, at $\mathcal{O}(\eta^2L_N^{-2})$. 

\subsection{Vector potentials}

\noindent
The $0i$-field equations at order $\mathcal{O}(\eta^3L_N^{-2})$ give
\bea
&\ & \nabla^2 {\mathbf B}^{(0,3)}_{i} + {2 }\left(a \dot{\Phi}^{(0,2)} + {\dot{a}}\Phi^{(0,2)}\right)_{,i} = 16 \pi a^2 \left( {\mathbf \rho}^{(0,0)} + {\mathbf \rho}^{(0,2)} +\mathbf{p}^{(0,0)} \right) {\mathbf v}^{(0,1)}_i \, .  \label{FINAL0i03} 
\eea
Although ${\mathbf B}^{(0,3)}_{i}$ is purely a divergenceless vector Eq. (\ref{FINAL0i03}) has a divergenceless vector and scalar part, which can be separated out with a derivative. At $\mathcal{O}(\eta^4L_N^{-2})$ the $0i$-field equations give
\bea
&&\nabla^2 \left({\mathbf B}^{(1,0)}_{i} +{\mathbf B}^{(1,2)}_{i}\right) +2 \left(a \left(\Phi^{(1,1)}-\Psi^{(1,0)}\right)\dot{•}+ \dot{a}\left(\Phi^{(1,1)}+ \Phi^{(1,0)} \right)   \right)_{,i}
 -2 \left( 2 {\dot{a}^2} + a {\ddot{a}} \right) {\mathbf B}^{(1,0)}_{i} - {\mathbf B}^{(1,0)}_{j}\Phi^{(0,2)}_{,ij} \label{FINAL0i04} \\[5pt]
&=& 8 \pi a^2  \left(2\left( \mathbf{\rho}^{(0,0)} + \mathbf{\rho}^{(0,2)} +\mathbf{p}^{(0,0)}  \right)\mathbf{v}^{(1,0)}_i +  2\mathbf{\rho}^{(1,1)}\mathbf{v}^{(0,1)}_i + \left(\mathbf{\rho}^{(0,0)} +\mathbf{\rho}^{(0,2)} + 3\mathbf{p}^{(0,0)}\right) \mathbf{B}^{(1,0)}_{i} \right) -2a^2 \mathbf{\Lambda} \mathbf{B}_{i}^{(1,0)}  \, , \; \; \; \; \; \; \nonumber
\eea
which can also be split into scalar and divergenceless vector part using a derivative. The reader may note that the quadratic term, which includes the lower-order potential $\Phi^{(0,2)}$, does not source the vector part of Eq. (\ref{FINAL0i04}).

\subsection{Higher-order scalar potentials}

\noindent
The $00$-field equation and the trace of the $ij$-field equation at $\mathcal{O}(\epsilon \eta L_N^{-2})$ gives 
\bea
&\ &   \nabla^2 \Phi^{(1,1)} = - 8 \pi a^2  {\mathbf \rho}^{(1,1)}\, , \label{FINAL0011} 
\eea
which implies
\be
\Phi^{(1,1)} = - \Psi^{(1,1)} \, . \label{condition11} 
\ee
Using the $00$-field equation at $\mathcal{O}(\eta^4L_N^{-2})$ is
\bea
&& \nabla^2 \left( \Phi^{(1,0)} + \frac{1}{2}\Phi^{(0,4)} + \Phi^{(1,2)}\right) + \left(\nabla \Phi^{(0,2)}\right)^2
+{3 a \dot{a}}\left(3\Phi^{(0,2)} + \Phi^{(1,0)} -2 \Psi^{(1,0)}\right)\dot{•} \nonumber \\[5pt]
&& + {3 a^2}\left(\Phi^{(0,2)}- \Psi^{(1,0)}\right)\, \ddot{•} 
 -\nabla^2\Phi^{(0,2)}\left( \Phi^{(0,2)} -\Psi^{(1,0)}\right) - \frac{1}{2} \Phi^{(0,2)}_{,ij}\mathbf{h}^{(1,0)}_{ij} \nonumber \\[5pt]
&=& -8 \pi a^2  \left[ \mathbf{\rho}^{(1,0)}  + \mathbf{\rho}^{(1,2)} + \frac{1}{2}\mathbf{\rho}^{(0,4)} - \left(\mathbf{\rho}^{(0,0)}+ \mathbf{\rho}^{(0,2)} +3\mathbf{p}^{(0,0)} \right) \left(\Phi^{(1,0)} + \Phi^{(0,2)} \right)  + 3 \left( \mathbf{p}^{(1,0)}+\mathbf{p}^{(1,2)} + \frac{1}{2}\mathbf{p}^{(0,4)} \right) \right] \nonumber \\
&& -16\pi a^2 \left( \mathbf{v}^{(0,1)}_i \right)^2 \left(\mathbf{\rho}^{(0,0)} + \mathbf{\rho}^{(0,2)} +\mathbf{p}^{(0,0)} \right) - 2a^2\mathbf{\Lambda} \left( \Phi^{(0,2)} +\Phi^{(1,0)} \right)  \, ,
\label{FINAL0004} 
\eea
while the trace of the $ij$-field equation at $\mathcal{O}(\eta^4L_N^{-2})$ gives
\bea
&& - 2 \nabla^2 \left(\Psi^{(1,0)} + \Psi^{(1,2)} + \frac{1}{2}\Psi^{(0,4)}\right) 
- 3\left(2 \dot{a}^2 + a \ddot{a}\right)\left(\Phi^{(1,0)} - \Psi^{(1,0)} + 2\Phi^{(0,2)}\right) 
+ 6\dot{a}a\left(\Psi^{(1,0)} -\Phi^{(0,2)} \right)\dot{}
 \nonumber \\[5pt]
&=& -16 \pi a^2 \left[   \mathbf{\rho}^{(1,0)} +  \frac{1}{2}\mathbf{\rho}^{(0,4)} + \mathbf{\rho}^{(1,2)} +\left(\mathbf{\rho}^{(0,0)} +\mathbf{\rho}^{(0,2)} +\mathbf{p}^{(0,0)}\right)\left( \mathbf{v}^{(0,1)}_i \right)^2  \right] \nonumber \\[5pt]
&& -4 \pi a^2 \left[ 2\Phi^{(0,2)} \left( \mathbf{\rho}^{(0,0)} +\mathbf{\rho}^{(0,2)} -3 \mathbf{p}^{(0,0)}  \right) - \left( \mathbf{\rho}^{(0,0)} + \mathbf{\rho}^{(0,2)}\right)\left( \Phi^{(1,0)} +3\Psi^{(1,0)} \right) + 3\mathbf{p}^{(0,0)} \left( \Psi^{(1,0)} - \Phi^{(1,0)} \right) \right]  \nonumber \\[5pt]
&& -a^2\mathbf{\Lambda} \left[ 4\Phi^{(0,2)} +\Phi^{(1,0)} -3\Psi^{(1,0)} \right]+ \mathcal{A}  \, , \label{FINALijtrace04}
\eea
where
\bea
\mathcal{A} &\equiv& \nabla^2 \Phi^{(0,2)}\left(3\Phi^{(0,2)} + \frac{1}{2}\Phi^{(1,0)} - \frac{5}{2}\Psi^{(1,0)}\right) + \frac{3}{2}\left(\nabla \Phi^{(0,2)}\right)^2 + \frac{1}{2}\Phi^{(0,2)}_{,ij}\mathbf{h}^{(1,0)}_{ij} \, . 
\label{FINALA}
\eea
These are all of the scalar equations that exist up to $\mathcal{O}(\eta^4L_N^{-2})$.

\subsection{Tensor potentials}

\noindent
The trace-free part of the $ij$-field equation at $\mathcal{O}(\epsilon \eta L_N^{-2})$ is
\bea
D_{ij}\left(\Phi^{(1,1)} + \Psi^{(1,1)} \right) - \frac{1}{2}\nabla^2 \mathbf{h}_{ij}^{(1,1)} =0 \, , \label{FINALij11}
\eea
which implies 
\be
\Phi^{(1,1)} = - \Psi^{(1,1)} \qquad {\rm and} \qquad \mathbf{h}^{(1,1)}_{ij}= 0 \, . \label{psiphihij11}
\ee
The reader may note that, unlike $\Psi^{(0,2)}$ and $\Phi^{(0,2)}$, the first of these conditions has already been given by the $00$-field equation and the trace of the $ij-$field equations (\ref{condition11}). Finally, the $\mathcal{O}(\eta^4L_N^{-2})$ part of the $ij$-field equation can be used to write
\bea
&& - D_{ij}\left(\Phi^{(1,0)} + \Phi^{(1,2)} + \frac{1}{2}\Phi^{(0,4)} + \Psi^{(1,0)} + \Psi^{(1,2)} + \frac{1}{2}\Psi^{(0,4)}\right) 
 + \frac{1}{2} \nabla^2 \left(\mathbf{h}_{ij}^{(1,0)} + \mathbf{h}_{ij}^{(1,2)} + \frac{1}{2}\mathbf{h}_{ij}^{(0,4)}\right) \nonumber \\[5pt]
&& +\frac{2}{a}\left[ a^2\left(\mathbf{B}_{(i,j)}^{(0,3)} + \mathbf{B}_{(i,j)}^{(1,0)}  \right) \right]\dot{} - \left(2\dot{a}^2 + a\ddot{a}\right)\mathbf{h}^{(1,0)}_{ij} 
 -\frac{3}{2} a\dot{a}\dot{\mathbf{h}}_{ij}^{(1,0)} - \frac{1}{2} a^2  \ddot{\mathbf{h}}_{ij}^{(1,0)}  \nonumber \\[5pt]
& =&  -4\pi a^2 \left[ \left( {\mathbf \rho}^{(0,0)} +{\mathbf \rho}^{(0,2)} -{\mathbf p}^{(0,0)} \right)  \mathbf{h}^{(1,0)}_{ij}+ 4 \left( {\mathbf \rho}^{(0,0)} +{\mathbf \rho}^{(0,2)} +{\mathbf p}^{(0,0)} \right) {\mathbf v}^{(0,1)}_{\langle i} {\mathbf v}^{(0,1)}_{j\rangle} \right] - a^2 \mathbf{\Lambda} \mathbf{h}^{(1,0)}_{ij} + \mathcal{B}_{ij}  \, , \; \; \; 
\label{FINALijtracefree04}
\eea
where 
\bea
\mathcal{B}_{ij} &\equiv& D_{ij}\Phi^{(0,2)}\left(2\Phi^{(0,2)} + \Phi^{(1,0)} -\Psi^{(1,0)}\right)  
+  \Phi^{(0,2)}_{,\langle i}\Phi^{(0,2)}_{, j \rangle} -  \Phi^{(0,2)}_{,k \langle i }\mathbf{h}^{(1,0)}_{j\rangle k} \, ,
\label{FINALBij}
\eea
and where we have used Eq. (\ref{psiphihij11}). Note that, unlike standard cosmological perturbation theory, these equations do not imply $\Phi^{(1,0)} = -\Psi^{(1,0)}$ or $\mathbf{h}_{ij}^{(1,0)} =0$, and that scalar, vector and tensor modes do not decouple at linear order in cosmological perturbations. This completes the full set of field equations in terms of our gauge-invariant variables, up to the order in perturbations that we wish to consider here.

\end{document}